\documentclass[final,3p,times]{elsarticle}

\usepackage{amssymb}
\usepackage{pifont}
\newcommand{\cmark}{\ding{51}} 
\newcommand{\xmark}{\ding{55}} 
\usepackage{makecell}
\usepackage{booktabs}
\usepackage{graphicx}
\usepackage{tikz}
\usetikzlibrary{positioning,arrows.meta}
\usepackage{subcaption}

\usepackage[font=small]{caption}

\usepackage{amsmath}
\usepackage{amsthm}
\usepackage{titlesec}
\titleformat{\paragraph}[runin]{\normalfont\bfseries}{\theparagraph}{1em}{}
\usepackage{enumitem}
\setlist{nosep}
\newtheorem{theorem}{Theorem}[section]   
\newtheorem{definition}[theorem]{Definition} 

\usepackage{tabularx}   
\usepackage{float}      
\usepackage{array}
\newcolumntype{P}[1]{>{\raggedright\arraybackslash}p{#1}}
\usepackage{lscape}
\usepackage{comment}
\usepackage{url}
\usepackage{titlesec}
\titlespacing*{\section}{0.3pt}{0.5ex}{0.3ex}
\titlespacing*{\subsection}{0.3pt}{0.4ex}{0.2ex}
\usepackage{setspace}
\journal{}

\begin{document}

\begin{frontmatter}



\title{Centrality Measures in Temporal Networks: A Critical and Comparative Survey}


\author[mymainaddress]{Aksa Urooj \corref{mycorrespondingauthor}}\ead{aksaurooj62@gmail.com}

\author[mymainaddress]{Iqra Altaf Gillani}
\ead{iqraaltaf@nitsri.ac.in}

\cortext[mycorrespondingauthor]{Corresponding author}

\address[mymainaddress]{Department of Information Technology, National Institute of Technology Hazratbal, Srinagar, Jammu and Kashmir, India}


\begin{abstract}
Temporal networks offer a suitable representation for complex systems in which interactions vary over time, such as communication, transportation, and social networks. Identifying influential nodes in such networks is more challenging than in static graphs because node importance depends not only on network structure but also on the timing and ordering of interactions. Although many temporal centrality measures have been proposed, the literature remains fragmented, with limited consensus on their comparative performance and applicability. This paper presents a critical and comparative survey of centrality measures in temporal networks. We review both temporal extensions of classical centrality metrics and measures specifically designed for temporal graphs, and propose a functional taxonomy that categorizes existing approaches according to the primary mechanism through which influence is quantified in temporal networks. The proposed taxonomy organizes temporal centrality measures into interaction-based, path-based, walk-based, spectral-based and robustness-based categories, providing a unified perspective on their underlying principles. In addition, we provide comparative insights to help select appropriate temporal centrality measures under different network characteristics and application settings. To complement the survey, we conduct experiments on multiple real-world temporal network datasets. The measures are evaluated through influence spreading experiments using epidemic diffusion models, ranking consistency analysis based on Kendall’s rank correlation, and runtime complexity analysis to assess computational efficiency and scalability. Finally, we highlight key open challenges and future research directions, including scalability for million-sized networks and the need for standardized evaluation frameworks.
\end{abstract}

\begin{keyword}
Temporal Networks, Centrality Measures, Influence Maximization, Node Importance, Online Social Networks
\end{keyword}

\end{frontmatter}

\section{Introduction}
Networks serve as an effective modelling framework for capturing interactions within complex systems\cite{boccaletti2006complex}. From the intricate web of neurons in the brain to the global infrastructure of the Internet, networks serve as a universal framework for modeling relationships and interactions\cite{holme2012temporal}. Traditionally, network analysis has concentrated on static graphs, in which relationships among entities are fixed, which can simplify management and reduce operational complexity. However, real-world systems are rarely static; interactions evolve with time\cite{li2013efficient}, giving rise to time-varying graphs called temporal networks. In such network structures, the interaction between any two nodes $x$ and $y$ from the vertex set $V$ is expressed as a quadruplet $(x,y,t, \delta t)$, where $t$ indicates the specific moment when nodes $x$ and $y$ engaged with one another and $\delta t$ is the time span of the interaction\cite{nicosia2013graph}. Temporal networks capture the dynamic nature of interactions, which makes them useful in modeling real-world systems where connections evolve and shift as time progresses. This dynamic nature means that both the graph structure and the interaction among nodes can change over time, requiring more sophisticated methods for analysis and representation. So, there is a tradeoff between getting a clear picture of the network and the complexity associated with it. Throwing away the time dimension reduces the complex nature of networks, but at the same time it loses its capability in applications like communication networks (phone calls are active at certain points)\cite{barabasi2005origin}\cite{stehle2010dynamical}, brain neural networks (where neurons are timely activated)\cite{holme2012temporal}, community evolution (tracking how communities form, merge, or dissolve over time)\cite{newman2004finding}\cite{berger2006framework}\cite{wang2011identifying}, disease spread (modeling how diseases spread through human contact over time)\cite{keeling2008modeling}\cite{anderson1991infectious}, information diffusion  (analyzing how information, rumors, or memes propagate through social media platforms over time)\cite{daley1965stochastic}\cite{medo2009adaptive}\cite{lu2011small}\cite{zhang2011closeness}, etc. As a result, analyzing temporal graphs requires specialized tools and approaches that balance accuracy with computational feasibility while capturing the temporally evolving behaviour of the system.\par
Within this broader field, centrality\cite{brede2012networks} metrics have been developed to understand the importance of nodes \cite{lu2016vital}. In static graphs, measures such as closeness \cite{sabidussi1966centrality}, degree \cite{gao2014ranking}, eigenvector\cite{bonacich1987power}\cite{bonacich1972factoring}, katz\cite{katz1953new}, betweenness\cite{freeman1977set}, pageRank \cite{brin1998anatomy}\cite{page1999pagerank}\cite{xing2004weighted} and coreness are widely used to identify influential or structurally critical nodes\cite{kitsak2010identification}. These metrics have guided applications in various fields such as identifying the right person for a viral marketing campaign\cite{kempe2003maximizing}, finding bottlenecks in traffic networks\cite{holme2003congestion}, infrastructure robustness\cite{holme2003congestion}, military communication systems\cite{kim2012temporal}, etc. However, when applied directly to temporal settings, static definitions often fail to account for the timing, ordering, and persistence of interactions. To overcome this limitation, researchers have introduced both temporal adaptations of classical centrality measures and novel centrality metrics designed specifically for temporal networks. Temporal adaptations modify existing definitions to incorporate time-dependent paths or dynamic snapshots, while temporal-specific measures focus explicitly on capturing evolving influence patterns. Together, these approaches form a rich and rapidly growing body of work that seeks to understand node importance under temporal constraints.\par
Despite this progress, there remains a lack of consolidated understanding of how different centrality measures perform in temporal contexts, their relative strengths and weaknesses, and their suitability for various application domains. Existing studies are often fragmented, focusing on specific measures or datasets, which makes it challenging for researchers and practitioners to identify the most appropriate metric for their problem. This survey bridges the gap between static centrality studies and the dynamic realities of temporal networks, offering a structured foundation for both researchers and practitioners.
\subsection{Comparison with Existing Surveys}
Existing surveys on temporal networks primarily focus on foundational modeling principles, temporal paths, and diffusion dynamics, with only partial coverage of node centrality measures. Some studies introduce specific temporal centrality formulations or frameworks for extending static metrics, while others address streaming or walk-based influence in isolation. In contrast, the present survey offers a comprehensive and critical comparison of centrality measures in temporal networks, encompassing both adaptations of classical metrics and measures specifically developed for temporal settings. By systematically analyzing their assumptions, computational properties, and application suitability, this work complements prior surveys and offers practical guidance for selecting appropriate centrality measures in evolving networks. To highlight the novelty and scope of our work, Table \ref{tab:survey_comparison} compares our survey with existing surveys on temporal networks and temporal centrality measures.
\begin{table}[H]
\small
\centering
\caption{Comparison of this survey with existing surveys on centrality measures in temporal networks}
\label{tab:survey_comparison}
\resizebox{\textwidth}{!}{
\begin{tabular}{lccccccc}
\toprule
\textbf{Survey} & \textbf{Year} & \textbf{Focus on Centrality} & \textbf{Multiple Centrality Types} & \textbf{Temporal Models} & \textbf{Comparative Analysis} & \textbf{Experimental Evaluation} & \textbf{Applicability}\\
\midrule
Holme \textit{et al.} \cite{holme2012temporal} & 2012 & \cmark & \xmark & \cmark & \xmark & \xmark &\xmark \\
Nicosia \textit{et al.} \cite{nicosia2013graph} & 2013 & \cmark & \cmark & \cmark & \xmark & \xmark & \xmark \\
Masuda \textit{et al.} \cite{masuda2016guide} & 2016 & \xmark & \xmark & \cmark & \xmark & \xmark &\xmark \\
Saxena \textit{et al.} \cite{saxena2020centrality} & 2020 & \xmark & \cmark & \xmark & \cmark & \xmark & \cmark \\
\midrule
\textbf{This Survey} & \textbf{2026} & \cmark & \cmark & \cmark & \cmark & \cmark &\cmark \\
\bottomrule
\end{tabular}
}
\vspace{2mm}
{\footnotesize \textbf{Note:} \cmark\ indicates the feature is addressed, while \xmark\ indicates it is not covered.}
\end{table}
\subsection{Our Contributions}  
This survey provides a comprehensive and critical analysis of centrality measures in temporal networks. 
The main contributions of this work are as follows:
\begin{itemize}
  \item We propose a functional taxonomy of temporal centrality measures based on the primary source of influence used to quantify node importance in evolving networks. The taxonomy groups existing approaches into Local Interaction-Based, Path-Based, Walk-Based Diffusion, Spectral, and Structural Robustness-Based categories.
  \item We present a comprehensive and up-to-date review of temporal centrality measures, encompassing both temporal adaptations of classical metrics and centrality measures specifically designed for evolving and streaming networks.
  \item We analyze real-world application domains to illustrate how different temporal centrality measures align with specific analytical objectives in social, healthcare, and financial networks.
  \item We provide an experimental comparison of several temporal centrality measures using real-world datasets, evaluating their effectiveness through influence spreading experiments, Kendall’s ranking correlation analysis and runtime complexity assessment to examine both influence identification capability and computational efficiency.
  \item We identify key open challenges and future research directions, highlighting unresolved issues related to scalability in streaming settings, and the integration of higher-order and multilayer temporal structures.
\end{itemize}
\subsection{Organization}
This survey bridges the gap between static centrality studies and the dynamic realities of temporal networks, offering a structured foundation for both researchers and practitioners. The rest of this paper is organized as follows. Section \ref{sec: Preliminaries} outlines the fundamental concepts necessary to comprehend the subsequent review. Section \ref{sec:taxonomy} presents a functional taxonomy of temporal centrality measures, categorizing existing approaches according to the primary mechanism through which influence is quantified in temporal networks. Section \ref{sec: CentralityM} explains one basic framework of each centrality measure that has been extended for temporal networks, along with its applications and limitations. Section \ref{sec: Centrality Temporal} explains various centrality methods developed particularly for temporal networks. Section \ref{sec:usecases} summarizes which temporal centrality measures are best suited for each network domain, the main analytical goal, and an example application. Section \ref{sec:exper_eval} presents the experimental evaluation of the compared temporal centrality measures using diffusion-based influence analysis, Kendall’s ranking correlation and runtime complexity assessment on real-world temporal network datasets. Section \ref{sec: oc} identifies key challenges such as scalability, real-time computation, and validation etc. It also suggests future directions, including unified evaluation frameworks, hybrid metrics, and domain-specific adaptations to advance temporal centrality research.
Finally, Section \ref{sec: conclusion} summarizes the paper by emphasizing its main findings and contributions.
\section{Preliminaries}  This section outlines the basic principles of temporal networks along with the diffusion models employed to evaluate the performance of the methods discussed in this survey.
\label{sec: Preliminaries}
\subsection{Definitions}
A \textit{temporal graph} can be viewed as a sequence of graphs arranged in chronological order \cite{tang2010analysing}. This dynamic graph $G^D_{(0,T)}$=$(V,E_{0,T})$ defined over a time interval $[0,T]$ comprises a fixed set of vertices $V$ and a set of temporal edges $E_{0,T}$. A temporal edge $(u,v)_{i,j} \in E_{0,T}$ connects vertices $u$ and $v$ during the time interval $[i,j]$, where 0 $\leq i \leq j \leq T$. In this network, the vertex set $V$ remains constant, while the set of edges $E_{0, T}$ evolves with time \cite{kim2012temporal}.

\begin{definition}(Cycle Basis \cite{lee2012qube})
It is a set of cycles in the graph that can be used to generate all other cycles in the graph using linear combinations. The total weight of a cycle basis is obtained by summing the weights of all its constituent cycles .
\end{definition} 
\begin{definition}(Minimum Cycle Basis (MCB) \cite{lee2012qube})
It is a cycle basis where the total weight of the cycles is the smallest possible.
\end{definition} 

\begin{definition}(Unstable Vertex Pair \cite{yen2013efficient})
For a given graph $G = (V, E)$, a vertex pair ${u, v} \in V$ is considered an unstable pair if the updated shortest path distance $|d'(u, v)|$, resulting from the addition or removal of an edge between $u$ and $v$, varies from the initial shortest path distance $|d(u, v)|$ .
\end{definition} 
\begin{definition}(r-Unstable Vertex Set \cite{yen2013efficient})
Given a BFS+ graph $G_r$ and its modified version $G'_r$, the $r$-unstable vertex set includes all vertices $v \in V$ whose levels change in $G'_r$. This set is defined as $V'_r = \{ v \mid |p'(r, v)| \neq |p(r, v)| \}$. If $V'_r = \emptyset$, then $r$ is considered a stable vertex, meaning that for every vertex $v \in V$, the condition $|p'(r, v)| = |p(r, v)|$ holds true.
\end{definition} 
\begin{definition}(Minimum Union Cycle(MUC) \cite{lee2012qube})
It is a single cycle that includes every edge of the graph and has the minimum possible weight. Since MUCs are mutually exclusive, each vertex belongs to exactly one MUC. We use MUC(v) to represent the MUC that includes the vertex $v$ .
\end{definition} 
\begin{definition}(Network Efficiency \cite{latora2001efficient})
It is the average of the inverse of the shortest path distance $d(u,v)$  and is given as: $E(G)=\frac{1}{n*(n-1)}\sum_{u \neq v \in G} \frac{1}{d(u,v)}$.
\end{definition} 
\subsection{Diffusion Models} To evaluate the spreading capability of nodes identified by various centrality measures, we employ two commonly used diffusion models for temporal networks: the Susceptible–Infected (SI) model and the Susceptible–Infected–Susceptible (SIS) model. These models represent different types of epidemic and information diffusion processes, enabling a comprehensive evaluation of influence in dynamic networks.
\paragraph{Susceptible–Infected (SI) Model}
In the SI diffusion model, nodes exist in one of two possible states: susceptible $(S)$ or infected $(I)$. During each time step, an infected node spreads the infection to its susceptible neighbors with transmission probability $\beta$. After infection, a node permanently remains in the infected state throughout the diffusion process.
\paragraph{Susceptible–Infected–Susceptible (SIS) Model}
In the SIS diffusion model, nodes can switch between susceptible $(S)$ and infected $(I)$ states. Infection spreads with probability $\beta$, while infected nodes recover with probability 
$\gamma$ and return to the susceptible state, allowing them to be infected again.
\section{Taxonomy of Temporal Centrality Measures}
\label{sec:taxonomy}
The growing availability of temporal network data has led to the development of a wide variety of centrality measures designed to identify influential nodes in evolving systems. Although these methods share the common objective of quantifying node importance, they differ substantially in the type of temporal information they exploit and the mechanism through which influence is assessed. Existing studies typically categorize temporal centrality measures according to their static counterparts, such as temporal degree, temporal closeness, or temporal betweenness. While such an organization is useful from a historical perspective, it does not adequately reveal the fundamental principles underlying different approaches.\par
To provide a more systematic understanding of the field, we propose a functional taxonomy of temporal centrality measures based on the primary source of influence used to determine node importance. Under this framework, temporal centrality measures are classified into five categories: (i) Local Interaction-Based Centralities, (ii) Path-Based Centralities, (iii) Walk-Based Diffusion Centralities, (iv) Spectral Centralities, and (v) Structural Robustness Centralities.
The proposed classification emphasizes how each measure interprets influence within a temporal network rather than focusing solely on its mathematical formulation. Such a perspective provides clearer guidance for selecting appropriate centrality measures in different application domains and highlights the trade-offs between computational efficiency, temporal awareness, and structural information. Table \ref{tab:taxonomy} provides a comparative overview of the proposed taxonomy.
\subsection{Local Interaction-Based Centralities}
Local interaction-based centralities determine node importance primarily through direct temporal interactions with neighboring nodes. These measures rely on information available within the immediate neighborhood of a node and typically avoid expensive global computations. The underlying assumption is that nodes participating in frequent or long-lasting interactions are more likely to influence their surroundings. Consequently, these methods are particularly effective in environments where local activity patterns strongly determine influence propagation. Representative examples include Time-Scale Degree Centrality (TSDC)\cite{uddin2011time}, Temporal Coreness Centrality\cite{li2013efficient} and other temporal degree variants\cite{elmezain2021temporal} that incorporate interaction duration and frequency. Unlike traditional degree centrality, these measures account for the temporal persistence of relationships, enabling a more realistic representation of influence in dynamic systems.\par
The main advantages of local interaction-based methods are their simplicity, scalability, and suitability for large-scale and streaming networks. However, because they focus only on immediate neighbors, they often fail to capture the broader structural position of a node within the network.
\subsection{Path-based Centralities}
Path-based centralities evaluate node importance through temporal paths connecting different parts of the network. In these approaches, a node becomes influential when it facilitates communication or information transfer between other nodes through time-respecting paths. Unlike local measures, path-based methods explicitly incorporate temporal causality by ensuring that information flows only along chronologically valid sequences of interactions. As a result, they provide a more realistic representation of communication and diffusion processes in temporal networks. Temporal closeness centrality\cite{tang2010analysing}, CENDY \cite{yen2013efficient}, QUBE \cite{lee2012qube}, Temporal Coverage Centrality (TCC), and Temporal Boundary Coverage Centrality (TBCC) belong to this category \cite{takaguchi2016coverage}. These methods quantify influence through temporal shortest paths, reachability, or the extent to which a node participates in efficient communication routes.\par
Path-based centralities are particularly useful in transportation systems, communication networks, and epidemic modeling, where information propagation depends heavily on temporal connectivity. Their primary limitation is computational complexity, as identifying temporal shortest paths often requires repeated network traversals across multiple time layers.
\subsection{Walk-Based Diffusion Centralities}
Walk-based diffusion centralities measure node importance by considering all possible temporal walks through the network rather than restricting attention to shortest paths. These methods are motivated by the observation that information, influence, or contagions do not necessarily follow optimal routes in real-world systems. In this category, influence emerges from a node’s ability to participate in and facilitate multiple temporally valid diffusion pathways. Both direct and indirect interactions contribute to the final centrality score, allowing these measures to capture complex spreading dynamics. Representative methods include Temporal Walk Centrality (TWC)\cite{oettershagen2022temporal}, Dynamic Centrality \cite{lerman2010centrality}, and TempoRank\cite{rocha2014random}. These approaches model influence propagation through temporally ordered walks, random walks, or diffusion processes that evolve across time.\par
As they consider a larger set of transmission pathways, walk-based methods often provide more realistic estimates of spreading capability than shortest-path-based approaches. Consequently, they are widely used in influence maximization, information dissemination, and epidemic spreading studies. However, the richer representation of diffusion dynamics generally comes at the cost of increased computational requirements.
\begin{table}[H]
\small
\centering
\caption{Proposed Taxonomy of Temporal Centrality Measures}
\label{tab:taxonomy}
\resizebox{\textwidth}{!}{
\begin{tabular}{p{3.5cm} p{3.5cm} p{4.5cm} p{4.5cm} p{4.5cm}}
\toprule
\textbf{Category} & \textbf{Representtaive Measures} & \textbf{Influence Source} & \textbf{Computational Cost} & \textbf{Typical Applications} \\
\midrule
Local Interaction-Based &
TSDC, Temporal Coreness &
Direct temporal interactions &
Low &
Contact Networks \\

Path-Based &
CENDY, QUBE, TCC, TBCC &
Temporal shortest paths and reachability &
High &
Communication Networks \\

Walk-Based &
TWC, Dynamic Centrality, TempoRank &
Temporal walks and diffusion dynamics &
Medium--High &
Influence maximization \\

Spectral-Based &
Supracentrality, IECM, Temporal PageRank, f-PageRank &
Eigenvector propagation&
High &
Social and citation networks \\

Structural Robustness &
Efficiency Centrality &
Network resilience &
Very High &
Infrastructure analysis \\

\bottomrule
\end{tabular}
}
\end{table}
\subsection{Spectral Centralities}
Spectral centralities determine node importance through recursive influence propagation mechanisms derived from matrix and eigenvector-based formulations. The central principle is that a node is important if it is connected to other important nodes, thereby extending local interactions into a global influence framework. Temporal spectral methods represent evolving networks using coupled layers or time-dependent adjacency structures. Centrality scores are then obtained from dominant eigenvectors or iterative propagation procedures that capture influence across both structural and temporal dimensions. This category includes Supracentrality \cite{taylor2017eigenvector}, Improved Eigenvector-based Centrality Measure (IECM)\cite{yin2018inter}, Temporal PageRank \cite{rozenshtein2016temporal}, Incremental PageRank \cite{desikan2005incremental}, and f-PageRank \cite{lv2019pagerank}. These methods provide a comprehensive view of network influence by integrating information from multiple temporal layers simultaneously.\par
Spectral centralities are particularly effective for identifying globally influential nodes in social, communication, and citation networks. Nevertheless, they often require substantial computational resources due to repeated matrix operations and the construction of large supra-adjacency structures.
\subsection{Structural Robustness Centralities}
Structural robustness centralities evaluate node importance according to its contribution to maintaining network cohesion, connectivity, or efficiency. Rather than focusing on information flow or diffusion dynamics, these measures assess the structural role of nodes within the evolving network. The key idea is that removing a highly important node should significantly degrade network functionality. Consequently, influence is quantified through the impact of node removal on structural properties such as connectivity or efficiency. Efficiency Centrality is a representative example of this category \cite{wang2017new}. Efficiency-based approaches evaluate how node removal affects the overall efficiency of communication across the network.\par
Structural robustness measures are particularly useful for infrastructure protection, network resilience analysis, and vulnerability assessment. However, because they focus primarily on structural stability, they may not accurately capture dynamic spreading processes that depend on temporal interaction patterns.
\subsection{Discussion}
The proposed taxonomy provides a unified framework for understanding the diverse landscape of temporal centrality measures. Each category captures a distinct aspect of node importance and is therefore suitable for different analytical objectives.\par
Local interaction-based centralities emphasize immediate activity and scalability. Path-based centralities focus on temporal reachability and communication efficiency. Walk-based diffusion centralities model realistic spreading dynamics. Spectral centralities capture global influence through recursive propagation mechanisms. Structural robustness centralities identify nodes that maintain network integrity and resilience. Since no single category is universally superior, the choice of a temporal centrality measure should depend on the application domain, network characteristics, and computational constraints. The proposed taxonomy not only clarifies the relationships among existing methods but also highlights opportunities for future research involving hybrid approaches that combine multiple sources of influence within a unified temporal framework.

\section{Classical Centrality Measures}
 \label{sec: CentralityM}
Identifying important nodes has long been a fundamental objective in the study of complex networks. Classical centrality measures, proposed for static network structures, have been widely applied across diverse domains, such as selecting optimal locations for service centers \cite{saxena2020centrality}, identifying target audiences for product promotion \cite{lewis2008tastes}\cite{borgatti2009network}, protecting key species in ecological systems \cite{jordan2008identifying}, and locating critical neurons in cortical networks \cite{bullmore2009complex}. These measures evaluate the significance of nodes using only the structural properties of the network, assuming that connections remain fixed over time. The following subsections discuss these standard centrality measures, which serve as the foundation for their temporal extensions examined later in this section.
\subsection{Degree Centrality (DC)}
It is one of the simplest and cost-effective centrality metrics \cite{bonacich1972factoring}. The basic idea is that nodes with more connections, or neighbors, contribute significantly to the spread of information than nodes with fewer connections. The DC of a node $u$ in static networks is defined using the following formula : $DC(u)=\frac{d_u}{n-1}$. Here, $d_u$ is the degree of node $u$.\par
\subsubsection{Degree Centrality Redefined for Temporal Networks}
\label{subsec:DegreeforTemporal}
Traditional DC focuses solely on the degree a node has, without taking into account the duration of those connections. However, a more detailed analysis reveals that the number of connections is not enough to accurately determine the most significant nodes in a network \cite{uddin2011time}, yet the length of these connections is equally important in identifying the important nodes. As an example, if a patient in a hospital interacts with five doctors, the patient’s degree centrality in the patient–doctor network becomes five, yet this value fails to reflect how long the treatment has been ongoing \cite {uddin2014new}. In this situation, the length of a relationship that develops and gets stronger with time \cite{ferreira2011evaluating} is crucial for her quick recovery. 
\begin{table}[H]
\small
\centering
\caption{Summary of works on Degree Centrality in Temporal Networks}
\label{table:degreecentrality}
\resizebox{\textwidth}{!}{
\begin{tabular}{p{3.5cm} p{3.5cm} p{4.5cm} p{4.5cm} p{4.5cm}}
\toprule
\textbf{Reference} & \textbf{Extension} & \textbf{Objective} & \textbf{Limitation} & \textbf{Application} \\
\midrule
Uddin \textit{et al.} \cite{uddin2011time} & New Direction in Degree Centrality Measure: Towards a Time-Variant Approach & To propose TSDC, a time-variant extension of DC that is based on both the presence and duration of ties in a network, overcoming the limitation of classical DC & Needs additional processing memory to store the adjacency matrices that keeps record of the number of edges between nodes and the time at which the edge is being added & Healthcare networks, inter-organizational collaboration, communication and social networks. \\



Kim \textit{et al.} \cite{kim2012temporal} & Temporal Node Centrality in Complex Networks & Provide a general method to extend static centralities (including degree) to time-ordered graphs & Time-ordered graph can become large, cost grows with sequence length & Human contact networks, mobile/ad-hoc systems \\

Elmezain \textit{et al.} \cite{elmezain2021temporal} & Temporal Degree-Degree and Closeness-Closeness: A New Centrality Metrics for Social Network Analysis & Propose refined temporal degree variants and hybrid metrics for social networks & Increased complexity and parameters, needs careful interpretation & Social network analysis, dynamic weighted networks \\
\bottomrule
\end{tabular}
}
\end{table}
Uddin \textit{et al.}\cite{uddin2011time} developed a time-sensitive model for measuring degree centrality, known as Time Scale Degree Centrality (TSDC). This approach incorporates number as well as the duration of connections between nodes in the network. 
Kim \textit{et al.} \cite{kim2012temporal} introduced the time-ordered graph model that reduces a temporal graph to a directed fixed graph, and used that construction to extend static centralities, such as degree, to the temporal case. Elmezain \textit{et al.} \cite{elmezain2021temporal} proposed temporal degree-degree, a refined temporal-degree variant that considers neighbor activity and other hybrid degree-like metrics for social networks. Table \ref{table:degreecentrality} presents a concise review of existing studies conducted on DC within temporal graphs.\par
DC is an easy-to-understand metric and is computable in time ($O(mT)$), where $m$ is the number of edges and $T$ is the number of snapshots. However, it is a basic measure as it overlooks the overall topology of the network. Moreover, measuring a node's degree is not an accurate way to identify the most significant node, because a node linked to a few important neighbors may spread information more effectively compared to a node connected to many less influential nodes.

\subsection{Coreness Centrality}
In their recent study, Kitsak \textit{et al.} \cite{kitsak2010identification} demonstrated that a node’s degree alone is not sufficient to determine its spreading influence. Instead, a more reliable indicator is the node’s location within the network. Coreness centrality assigns each node a core number that shows which layer it belongs to. A higher number means the node is part of a more tightly connected, central group in the network \cite{garas2012k}. Seidman \cite{seidman1983network} introduced the concept of k-cores as a powerful and computationally efficient tool to find influential nodes in a network. A k-core is a group of connected nodes where each node is linked to at least $k$ other nodes within that group. The coreness of a node $v$ is given as $C_S(v) = i$ if it is part of a maximal connected subgraph $H(V_1, E_1)$ such that $\forall v \in V_1$, the condition $k_v \geq i$ holds.
. $k_v$ is degree of node $v$ in induced subgraph $H$ \cite{saxena2020centrality}.\par
\subsubsection{Coreness Centrality redefined for Temporal Networks} 
In many real-world networks, the network typically evolves over time \cite{holme2012temporal} \cite{holme2015modern} \cite{saramaki2013temporal}, and it is computationally impractical to recalculate the core value of every node following each edge addition or removal. The first incremental methods for $k$-core decomposition in streaming networks were introduced by Sariyüce \textit{et al.}\cite{sariyuce2013streaming}, demonstrating that their approach achieves a speed-up of up to a million times compared to the traditional k-shell method on a network with 16 million nodes. 
\begin{table}[H]
\small
\centering
\caption{Summary of works on Coreness Centrality in Temporal Networks}
\label{tab:corecentrality}
\resizebox{\textwidth}{!}{
\begin{tabular}{p{3.5cm} p{4.5cm} p{5.5cm} p{4.5cm} p{4cm}}
\toprule
\textbf{Reference} &
\textbf{Extension} &
\textbf{Objective} &
\textbf{Limitation} &
\textbf{Application} \\
\midrule

Sariyuce \textit{et al.}~\cite{sariyuce2013streaming} &
Streaming Algorithms for k-core Decomposition &
Presented the first incremental approach to $k$-core decomposition tailored for streaming graph datasets. &
Edge insertions between vertices with large k-core values have large execution times &
Community detection, protein function prediction \\
Li \textit{et al.}~\cite{li2013efficient} &
Efficient Core Maintenance in Large Dynamic Graphs &
Proposing an Efficient Approach to Update and Maintain Node Core Numbers in Evolving Graphs &
Low Performance with low core number edges &
Monitoring dynamics of cohesive subgroups \\
Galimberti \textit{et al.}~\cite{galimberti2020span} &
Span-core Decomposition for Temporal Networks: Algorithms and Applications & Introduced span-core decomposition to identify dense subgraphs with temporal span in temporal networks. &
Computational complexity increases quadratically with the number of time intervals & Epidemic modeling, community detection, anomaly detection \\
Jakma \textit{et al.}~\cite{jakma2012distributed} &
Distributed k-Core Decomposition of Dynamic Graphs &
To design and present the first published continuous, distributed k-core decomposition algorithm for dynamic graphs & Complex to handle link deletions &
Network visualization, analyzing the topological structure of large networks \\
\bottomrule
\end{tabular}}
\end{table}
Li \textit{et al.} \cite{li2013efficient} developed a method to efficiently determine and update the core values of affected nodes following graph modifications. The algorithm iteratively removes nodes, beginning with the lowest degree, while initially assigning color 0 to all nodes. When an edge $Q-S$ gets added, the core number updates are determined using the standard k-core decomposition method. Galimberti \cite{galimberti2020span} introduced the concept of temporal core decomposition, in which each core is characterized by two attributes: its coreness, measuring the density of connections, and its span, representing a temporal interval. Jakma \textit{et al.} \cite{jakma2012distributed} introduced the first distributed approach capable of continuously computing the shell-index in dynamic graphs, overcoming the shortcomings of conventional centralized approaches that require global information about the graph. Table \ref{tab:corecentrality} provides an extensive summary of different approaches and algorithms associated with coreness centrality in temporal networks. 
\par 
Coreness Centrality can be computed by performing k-core decomposition on each temporal snapshot with a complexity of $O(T(n+m))$, where $T$ is the number of time snapshots. This measure often assigns the same core number to many nodes, lacking granularity. Furthermore, since nodes are classified into k-shells, the nodes within the same shell exhibit similar spreading capabilities.

\subsection{Closeness centrality}
Coreness centrality identifies nodes that belong to tightly interconnected subgraphs, revealing influential nodes within cohesive regions of the network. Although it effectively captures local structural prominence, it overlooks the global perspective by not considering how easily a node can reach others across the entire network. To address this limitation, closeness centrality (CC) evaluates node importance based on the average shortest-path distance to all other nodes, thereby reflecting its reachability and the effectiveness of information spread. It calculates how near a node is to every other node in the network. 
Alex Bavelas \cite{bavelas1950communication} defined it as the reciprocal of the farness: $CC(v)= \sum_{u \neq v} \frac{1}{d(v,u)}$. Here, $d(v,u)$ is the shortest path distance between nodes $v$ and $u$. The $CC$ of a node $v$ in a non-evolving network is given as, $CC(v)=\frac{n-1}{\sum_{\forall{u},v\neq u}d(v,u)}$ \cite{freeman2002centrality} . 
\subsubsection{Closeness Centrality redefined for Temporal Networks }
A static aggregated network typically contains more connections than those found in real-world scenarios, where the structure evolves continuously with the creation and removal of connections \cite{kim2012temporal}. As a result of this, the paths between nodes are overestimated, and the geodesic distance is underestimated \cite{tang2009temporal}. Moreover, the timing of interactions is not considered, so the estimates generated by static closeness centrality are often inaccurate. To fix these problems, Tang \textit{et al.} \cite{tang2010analysing} introduced the concept of temporal closeness centrality, which incorporates time information by using temporal shortest paths. The temporal CC can be expressed as :$CC(v)=\frac{1}{W(n-1)}\sum_{v \neq u \in V} d(v,u)$. Here, $W$ is the total number of temporal graphs.\par
Yen introduced the CENDY algorithm (Closeness centrality and avErage path leNgth in DYnamic networks) to efficiently update CC in dynamic networks upon the addition or removal of edges \cite{yen2013efficient}. We first compute the initial CC of all nodes, and then, after the edge insertion, we detect the unstable vertices (those whose distances to other nodes have changed due to the new edge). Finally, we calculate the updated closeness centrality of affected vertices.
Sariyuce \textit{et al.} introduced a technique for updating closeness centrality by utilizing level difference information obtained from a breadth-first traversal \cite{sariyuce2013incremental}. The proposed algorithm proved to bring a 99 times speedup in large networks containing over 500,000 edges. 
Elmezain \textit{et al.} \cite{elmezain2021temporal} proposed temporal closeness-closeness, a refined temporal-closeness variant that considers neighbor activity and other hybrid closeness-like metrics for social networks.
Oettershagen \textit{et al.} \cite{oettershagen2022computing} focused on the computation of top-k temporal closeness centrality using harmonic distance. Instead of recalculating full closeness scores, their algorithms extract only the most central nodes, achieving strong efficiency gains. The works done in the direction of CC in temporal settings are given in Table \ref{table:closecentrality}.
\begin{table}[H]
\small
\centering
\caption{Summary of works on Closeness Centrality in Temporal Networks}
\label{table:closecentrality}
\resizebox{\textwidth}{!}{
\begin{tabular}{p{3.5cm} p{3.5cm} p{4.5cm} p{4.5cm} p{4.5cm}}
\toprule
\textbf{Reference} & \textbf{Extension} & \textbf{Objective} & \textbf{Limitation} & \textbf{Application} \\
\midrule
Yen \textit{et al.} \cite{yen2013efficient} & An Efficient Approach to Updating Closeness Centrality and Average Path Length in Dynamic Networks & To design an efficient algorithm (CENDY) for updating closeness centrality of vertices and the average path length in dynamic networks when edges are inserted or deleted. & Not directly applicable to weighted or directed networks. & Social networks, biological and sensor networks. \\

Sariyuce \textit{et al.} \cite{sariyuce2013streaming} & Incremental Algorithms for Network Management and Analysis based on Closeness Centrality & To design incremental algorithms that efficiently update closeness centrality values when a network undergoes edge insertions or deletions & Algorithms are developed for unweighted and undirected graphs & Infrastructure and power grids, transportation/airline networks \\


Elmezain \textit{et al.} \cite{elmezain2021temporal} & Temporal Degree-Degree and Closeness-Closeness: A New Centrality Metrics for Social Network Analysis & Aims to capture both how close a node is over time and closeness of neighbors. & Hybrid measures may be harder to interpret. & Social network analysis, weighted interaction networks \\

Oettershagen \textit{et al.} \cite{oettershagen2022computing} & Computing Top-k Temporal Closeness in Temporal Networks & Develop algorithms to compute exact top‐k nodes by temporal closeness & Focuses on top‐k rather than full centrality & Spreaders in epidemic models, network monitoring. \\

\bottomrule
\end{tabular}}
\end{table}
CC requires calculating time-respecting shortest paths from each node to all others, leading to a complexity of $O(Tnm)$, where $T$ is the number of snapshots. Although it considers the global network structure, it still has a few limitations. A key drawback is its inapplicability to networks with disconnected components, as nodes in separate components have an undefined distance between them. Due to this, the method becomes inefficient for handling large-scale networks effectively.
\subsection{Betweenness Centrality}
In complex network analysis, it is vital to determine prominent nodes that may represent humans in a society, elements in biological networks \cite{sporns2007identification} \cite{jeong2001lethality}, people in organizations who mediate the communication\cite{teng2016collective} among employees, or junctions in transportation networks. Although closeness centrality captures global accessibility, it neglects the intermediary role of nodes that connect different network regions. This limitation motivates the use of betweenness centrality (BC), which highlights nodes acting as essential bridges in the network. This centrality measure was introduced by Freeman \cite{freeman2002centrality} to determine individuals who play a central role in mediating communication paths as they evolve with time. 
For a given node $v$, $BC$ represents the ratio of shortest paths connecting all node pairs that pass through $v$, and is mathematically expressed as: $BC(v)=\sum_{s \neq t \neq v \in V }\frac{d_v(s,t)}{d(s,t)}$. Here, ${d_{v}(s,t)}$ indicates the number of shortest routes from $s$ to $t$ that pass through $v$. The relative centrality of $v$ in a graph may be expressed as a ratio : $BC(v)=\frac{1}{(n-1)(n-2)}\sum_{s \neq t \neq v \in V }\frac{d_{v}(s,t)}{d(s,t)}$ \cite{freeman1977set}.
\subsubsection{Betweenness Centrality redefined for Temporal Networks }Traditional BC metrics rely only on the total count of shortest paths passing through a node, without accounting for the time a node stores a message prior to forwarding it to the subsequent node in the path.
Moreover, the computational cost of BC is intensive compared to other centrality measures, with a time complexity of $O(Tn^3)$, as it involves counting all shortest paths that traverse through each node for every possible pair of nodes in the network. Therefore, recalculating BC for all vertices from scratch is not practical, as even a single insertion or deletion of an edge can alter numerous shortest paths, which consequently affects the BC of many nodes in the graph. Consequently, several approaches have been developed to update its value within dynamic networks. \par
The first method for efficiently recomputing BC in dynamic networks was developed by Kas \textit{et al.} \cite{kas2013incremental} by reusing prior shortest-path computations and updating only affected nodes and edges. Later, Lee \textit{et al.} \cite{lee2012qube} introduced the QUBE framework, that confines updates to a candidate vertex set called the Minimum Union Cycle (MUC). When edges are added or removed, only the affected MUCs are updated, significantly reducing computation. 

\begin{table}[H]
\small
\centering
\caption{Summary of works on Betweenness Centrality in Temporal Networks}
\label{table:betweennesscentrality}
\resizebox{\textwidth}{!}{
\begin{tabular}{p{3.5cm} p{3.5cm} p{4.5cm} p{4.5cm} p{4.5cm}}
\toprule
\textbf{Reference} & \textbf{Extension} & \textbf{Objective} & \textbf{Limitation} & \textbf{Application} \\
\midrule
Kas \textit{et al.} \cite{kas2013incremental} & Incremental Algorithm for Updating Betweenness Centrality in Dynamically Growing Networks & To propose update algorithms for dynamic graphs that eliminate the need for complete recomputation of BC & Handles only edge additions, not deletions & Dynamic collaboration and co-authorship networks. \\

Lee \textit{et al.} \cite{lee2012qube} & QUBE: a Quick algorithm for Updating BEtweenness centrality & To address the challenge of updating BC in dynamic networks & Does not reflect global efficiency losses in the network. & Dynamic social networks, communication networks \\

Kourtellis \textit{et al.} \cite{kourtellis2015scalable} & Scalable Online Betweenness Centrality in Evolving Graphs & To design a scalable, online algorithm that updates BC efficiently & Focuses on structural evolution, not explicit temporal ordering of events. & Real-time web or traffic networks, evolving collaboration networks \\


Cruciani \textit{et al.} \cite{cruciani2024mantra} & MANTRA: Temporal Betweenness Centrality Approximation through Sampling & Efficient approximation of temporal BC in dynamic networks & Limited to betweenness-type measures; may underperform on networks with highly irregular temporal dynamics or dense time layers & Transportation and communication network monitoring \\
\bottomrule
\end{tabular}}
\end{table}
Green \textit{et al.} \cite{green2012fast} developed a method for efficiently modifying BC values in evolving graphs, which uses quadratic storage space to reduce computation time significantly, making it impractical for use in large-scale networks. Kourtellis \textit{et al.}\cite{kourtellis2015scalable} expanded upon the work in \cite{green2012fast} to support fully dynamic updates, achieving improvements in both space and time efficiency. This stream-based update mechanism maintains approximate betweenness scores using incremental shortest-path updates and pruning techniques. The algorithm supports both insertions and deletions in dynamic graphs. 
Cruciani \textit{et al.} \cite{cruciani2024mantra} introduced MANTRA, a scalable method to approximate temporal betweenness with provable error bounds while drastically reducing computation time. MANTRA provides accurate temporal betweenness approximations with significant speed-up and minimal accuracy loss compared to exact temporal betweenness methods. An overview of various techniques developed for BC is given in Table \ref{table:betweennesscentrality}.\par
While BC accounts for the global structure of a graph and can be utilized even in graphs with disconnected components, it possesses certain drawbacks. For instance, in many networks, a large number of nodes do not appear on the shortest paths between any two other nodes, leading to a centrality score of zero. Moreover, it focuses solely on the number of shortest paths traversing a node, assuming that information consistently travels along these paths. Nevertheless, this assumption is often unrealistic in practical cases, because information or infections can also diffuse over longer paths with some chance. So, Newman \cite{newman2005measure} introduced a betweenness measure that relaxes this assumption by considering all random walks between nodes, rather than limiting to the shortest ones.
\subsection{Eigenvector Centrality}
One limitation of the degree measure is that it assigns equal weight to all neighbors of a node when determining its importance. Eigenvector centrality (EC) generalizes DC and considers not just how many links a person or point has, but also how well-connected their neighbors are by repeatedly adding up the eigen vector values of the node’s adjacent vertices. The EC of node v can be expressed as \cite{brohl2019centrality} : $EC(v) =\sum_{u}A_{vu} EC(u)
$. Here, $A_{vu}$ is the adjacency matrix of the graph and $EC(u)$ is the EC of the neighbour of $v$.\par

\subsubsection{Eigen-vector Centrality redefined for Temporal Networks }
With the growing availability of dynamic network data, understanding how the importance of nodes evolves becomes increasingly crucial. So, Taylor \textit{et al.}  \cite{taylor2017eigenvector}  presented a general framework for applying EC methods to evolving graphs. The authors propose a method that models a dynamic network as a sequence of interconnected network layers, each representing a different time window. Specifically, they construct what is called a supracentrality matrix, a large matrix that incorporates both within-layer (intralayer) connections and between-layer (interlayer) links.
The supracentrality matrix’s dominant eigenvector is then used to determine the centrality of each node at every time layer. 
Yin \textit{et al.} \cite{yin2018inter} introduced an improved Eigenvector-based Centrality Measure (IECM) that replaces the fixed interlayer coupling parameter $\omega$ in the ECM with interlayer similarity based on node-level measures (e.g. Common Neighbors, Jaccard Index). The IECM method provides a more accurate and flexible way to compute centrality in evolving networks by incorporating adaptive inter-layer similarity, addressing the limitations of fixed coupling parameters in ECM. Flores \textit{et al.} \cite{flores2018eigenvector} developed an analytical model for eigenvector-like scores in continuous-time temporal networks and proved properties about existence/uniqueness and evolution. It treats continuous-time evolution and provides theoretical guarantees. A summary of these studies is provided in Table \ref{table:eigenvectorcentrality}.\par
\begin{table}[H]
\small
\centering
\caption{Summary of works on Eigenvector Centrality in Temporal Networks}
\label{table:eigenvectorcentrality}
\resizebox{\textwidth}{!}{
\begin{tabular}{p{3.5cm} p{3.5cm} p{4.5cm} p{4.5cm} p{4.5cm}}
\toprule
\textbf{Reference} & \textbf{Extension} & \textbf{Objective} & \textbf{Limitation} & \textbf{Application} \\
\midrule
Taylor \textit{et al.} \cite{taylor2017eigenvector} & Eigenvector-based Centrality measures for Temporal Networks & Provide a principled, general framework to extend any eigenvector-based centrality to temporal networks via a supracentrality matrix. & Requires choices for interlayer coupling and layer centrality matrices; supra-matrix of size NT can be large for long time series & Social, citation, co-appearance networks \\

Taylor \textit{et al.} \cite{taylor1904tunable} & Tunable Eigenvector-based Centralities for Multiplex and Temporal Networks & Generalize and tune supracentrality approach to multiplex and temporally coupled layers and analyze dependence on coupling topology and strength & Need to estimate interlayer coupling; computational cost for very large NT & Multilayer temporal problems \\

Yin \textit{et al.} \cite{yin2018inter} & Inter-layer Similarity-based Eigenvector Centrality Measures for Temporal Networks & Developed an eigenvector-based temporal centrality utilizing interlayer similarity to better track node influence dynamics than standard eigenvector centrality & Only considers similarity between consecutive layers; may miss long-term influence trends & Dynamic social networks, communication networks \\

Flores \textit{et al.} \cite{flores2018eigenvector} & On Eigenvector-like Centralities for Temporal Networks: Discrete vs. Continuous Time Scales & Propose continuous-time / analytical modeling of eigenvector-like centralities for temporal networks evolving in continuous time & More mathematical, less applied examples; continuous-time data are harder to obtain and model & Theoretical investigations, continuous-time systems (communication traces, sensor streams) \\

\bottomrule
\end{tabular}}
\end{table}
EC is computed iteratively for each snapshot, with per-snapshot complexity of $O(k(n+m))$, where $k$ is the number of iterations. It is not well-suited for directed acyclic networks that aren't strongly connected. In these networks (like citation networks), if a network isn’t strongly connected, nodes outside the main strongly connected component might have zero centrality scores, even if they are important. This can lead to an unfair representation of node importance in directed networks. Additionally, EC primarily focuses on overall network influence, and it may underestimate the importance of nodes that are influential within smaller clusters but not well-connected to the broader network. Also, the power-iteration method used to calculate EC can fail to converge or converge slowly in some cases. This can lead to inaccurate or incomplete results, especially for bipartite graphs. The algorithm’s iterative nature can cause significant computational overhead in large-scale networks.

\subsection{PageRank Centrality}
As the Web continues to expand rapidly, delivering high-quality, relevant pages in response to user queries becomes more challenging. This is partly because some web pages lack clear, descriptive content, and some links serve only navigational functions rather than indicating content relevance. To address these problems, several algorithms have been proposed. PageRank (PR) is a popular method created to rank web pages in search results \cite{langville2011google}\cite{gleich2015pagerank}. It’s named after one of Google’s founders, Larry Page \cite{page1999pagerank}, who helped develop the idea. It is a global centrality metric used to compute the universal rank of all web pages based on their position within the web, without considering the actual content of the pages. It is mostly applicable in search engines, browsing, and social network analysis. Each page contains a certain number of forward links (outgoing links) and backward links (incoming links). A page gets a high ranking if it is connected to other pages that are also highly ranked. The rank score of a web page $i$ is given by: $ R_i=c \sum_{ j\in B_i} \frac{R(j)}{N_j}$. Here, $B_i$ is the group of pages that point to $i$, $N_j$ is the total number of outgoing links from $j$, and $c$ is a normalization constant used to ensure that the overall rank across all web pages remains consistent.\par 
\subsubsection{PageRank Centrality redefined for Temporal Networks }
The structure of the web evolves dynamically as new pages are created and existing ones are deleted, and links between them are added or removed. Recomputing the full PR from scratch every time changes happen is very inefficient. So, Desikan \textit{et al.} \cite{desikan2005incremental} introduced an algorithm to update PR only for the parts of the graph that have changed, saving time and computing resources. The incremental PR algorithm partitions the updated graph into two parts, $P$ and $Q$. $P$ is the unchanged part of the graph since the last update, and $Q$ includes all the changed or affected nodes (either because their links changed or they were influenced by changed nodes). The authors use the fact that PR follows a first-order Markov model, which means each node's importance depends only on the nodes that directly point to it.
Rozenshtein \textit{et al.} \cite{rozenshtein2016temporal} extended the classic PR algorithm to temporal networks by incorporating time-stamped edges. The method ranks nodes based on their evolving influence over time, capturing temporal dynamics ignored by static PR.
\begin{table}[H]
\small
\centering
\caption{Summary of works on PageRank Centrality in Temporal Networks}
\label{table:pagerankcentrality}
\resizebox{\textwidth}{!}{
\begin{tabular}{p{3.5cm} p{3.5cm} p{4.5cm} p{4.5cm} p{4.5cm}}
\toprule
\textbf{Reference} & \textbf{Extension} & \textbf{Objective} & \textbf{Limitation} & \textbf{Application} \\
\midrule
Desikan \textit{et al.} \cite{desikan2005incremental} & Incremental Page Rank Computation on Evolving Graphs & Proposed incremental PR computation for evolving graphs, updating only affected nodes to reduce computation time. & Assumes slow changes in the graph; less effective if the graph changes rapidly & Web search engines, social networks, recommendation systems, web mining. \\

Rozenshtein \textit{et al.} \cite{rozenshtein2016temporal} & Temporal PageRank & Introduced temporal PR, extending the classic PR to temporal networks by considering time-stamped edges. & Assumes uniform edge arrival times; may not capture real-time dynamics accurately. & Social media analysis, temporal citation networks. \\

Lv \textit{et al.} \cite{lv2019pagerank} & PageRank Centrality for Temporal Networks & Proposed f-PageRank, a centrality measure that ranks both nodes and temporal layers simultaneously in evolving networks. & Computationally intensive for large-scale networks; requires solving eigenvectors of multi-homogeneous maps. & Dynamic social networks, temporal communication systems. \\

Aleja \textit{et al.} \cite{aleja2024time} & Time-dependent Personalized PageRank for Temporal Networks: Discrete and Continuous Scales & Developed a time-dependent personalized PR for temporal networks, incorporating time-varying personalization vectors. & Complexity increases with the number of time layers; may face scalability issues. & Personalized recommendation systems, evolving social networks. \\

\bottomrule
\end{tabular}}
\end{table}
 Lv \textit{et al.} \cite{lv2019pagerank} proposed f-PageRank, which simultaneously ranks both nodes and time layers, to reflect the interdependence between node significance and temporal dynamics. While it improves ranking accuracy, it is computationally intensive for large-scale networks. The research carried out by Aleja \textit{et al.} \cite{aleja2024time} introduces a time-dependent personalized PR that adapts the personalization vector over time. It effectively identifies influential nodes in evolving networks, but may face scalability issues with many time layers. In this context, several studies have been conducted, which are summarized in the following table (see Table \ref{table:pagerankcentrality}).\par 
Temporal PR is based on iterative propagation across time-ordered graphs, with an overall complexity of $O(Tk(n+m))$, where T is the number of temporal snapshots and k is the number of iterations. Although the PR algorithm has been effectively implemented by Google, a limitation remains: in real-world web pages, certain links may carry more significance than others. This method also tends to favor older pages, since newer ones haven’t had enough time to collect links and boost their ranking \cite{baeza2002web}. Multiple approaches have been introduced to address the recency bias inherent in PR, as discussed in \cite{berberich2005time} \cite{dong2010towards} \cite{yu2004temporal}.
\par
\section{Centrality Measures in Temporal Networks}
\label{sec: Centrality Temporal} 
Static centrality measures and their temporal extensions often fail to capture the true dynamics of evolving systems, as they typically aggregate time-varying connections or treat temporal snapshots independently. Such approaches ignore the causality, sequencing, and timing of interactions, factors that critically influence how information or influence actually propagates through a network. These limitations have motivated the development of centrality measures designed specifically for temporal networks, which move beyond aggregated views and instead exploit the temporal dimension to evaluate influence, accessibility, and control in evolving systems. 
Various works done in the direction of centrality metrics developed specifically for temporal networks are summarized in Table \ref{table:temporalcentrality}.

\subsection{Efficiency Centrality}
Efficiency centrality (EffC) is introduced as a method to evaluate the significance of an individual node across the entire network. It finds important nodes by removing them one at a time and measuring the resulting decrease in the overall network efficiency. The effect of node removal varies depending on its importance. For example,  eliminating a key node can significantly alter network structure and efficiency, such as modifying the shortest paths between nodes or affecting overall connectivity. Conversely, removing an isolated or less significant node produces minimal impact on the overall network structure.
The EffC of node $u$ is given as \cite{wang2017new}: 
$C_{Effc}(u)=\frac{E(G)-E(G')}{E(G)}, u \in V 
$. Here, $E(G')$ gives the EffC of the network $G'$ when node $u$ is removed. This method is more effective because it considers the importance of each node to the overall efficiency of complex networks. Moreover, unlike CC, EffC can be used in any kind of network, including directed ones. 

\subsection{Temporal Walk Centrality}
Oettershagen \textit{et al.} \cite{oettershagen2022temporal} introduced temporal walk centrality (TWC), a measure that evaluates a node’s ability to receive and transmit information over time. Unlike traditional measures, it incorporates the duration and sequence of interactions, capturing how timing influences information flow. It accounts for time-based paths through each node, assigning weights based on path length and temporal spread. The TWC of a node $u$ is defined by: $TWC(u)=\sum_{t_1,t_2\in T(G),t_1\leq t_2}(W_{in} (u,t_1).W_{out}(u,t_2) .\phi_m (t_1,t_2))$. Here, $W_{in} (u,t_1)$ and $W_{out}(u,t_2)$ represent the incoming and outgoing temporal walks, weighted by functions $\phi_{in}$ and $\phi_{out}$, respectively. The time-dependent weight function $\phi_m$ accounts for the duration between receiving and forwarding information, and can be defined by either walk length $(\phi(t_1,t_2)=\alpha, 0<\alpha<1)$ or waiting time $\left(\phi(t_1,t_2)=\frac{1}{1+t_2-t_1}\right)$. A combined formulation, $\phi(t_1,t_2)=\frac{\alpha}{1+t_2-t_1}$, captures both structural and temporal effects, with weights decreasing for longer paths or delays.
Unlike other measures (temporal betweenness, Katz, PageRank, closeness, and their static variants), TWC uniquely identifies nodes that cannot pass information and better reflects their role in dissemination processes. 

\subsection{Dynamic Centrality}

Real-world networks evolve continuously as nodes and edges change over time, making static graph representations inadequate for accurately estimating node importance \cite{kim2012temporal}. To address this limitation, Lerman \cite{lerman2010centrality} proposed a dynamic centrality measure based on time-respecting paths, extending $\alpha$-centrality \cite{bonacich2001eigenvector} to temporal networks represented as a sequence of snapshots $\triangle_{1,n}=\{t_1,\dots,t_n\}$ with adjacency matrices $A(t_i)$.
Two variants are introduced:
\begin{itemize}
    \item{Memoryless model:} assumes information can only propagate to the next time step. Dynamic centrality is computed using temporally ordered products of adjacency matrices weighted by parameters $\beta$ and $\alpha$, representing information initiation and propagation probabilities.
    
    \item{Memory-based model:} incorporates historical interactions through a memory decay parameter $\gamma$. Past connections are aggregated with exponentially decreasing weights, enabling the model to identify persistent influential nodes over time.
\end{itemize}
Experiments on toy examples and large citation networks demonstrate that dynamic centrality produces rankings substantially different from static centrality and PageRank, while better capturing temporally constrained information flow and hidden influential nodes.
\subsection{Dynamic - Sensitive Centrality}
Huang \textit{et al.}\cite{huang2017dynamic} proposed an extension of the Dynamic-Sensitive Centrality (DSC) \cite{liu2016locating} framework to temporal networks by combining nodal temporal dynamics with network topology. The proposed centrality is derived from the SIR epidemic model, using a discrete-time Markov chain to model infection probabilities across time layers. By representing the evolving network as a sequence of multilayer adjacency matrices, the method quantifies each node’s cumulative influence on spreading processes over time. This temporal extension effectively captures how time-order and dynamic interactions impact the propagation potential of nodes, outperforming traditional static and temporal degree, closeness, and betweenness-based centralities in identifying key spreaders in both real-world and artificially generated networks.
Mathematically, the influence of node $i$ at time $t$, denoted by $S_i(t)$, is obtained through matrix operations involving the temporal adjacency matrices and parameters related to infection and recovery rates. This method accounts for how the temporal order and the dynamic evolution of links affect the spreading potential of each node, providing a principled way to rank nodes by their dynamic centrality.
\subsection{Temporal Coverage Centrality (TCC)}
It quantifies the proportion of all vertex pairs in a network whose fastest temporal path includes the given temporal node \cite{takaguchi2016coverage}. This metric relies on the principle of coverage: a temporal vertex $v=(v,\tau)$ is said to cover a pair of vertices $(u,w)$ if passing through $v$ does not increase the travel time for the fastest possible path between $u$ and $w$. Specifically, the pair $(u,w)$ is covered by $v$ if two conditions hold when considering the latest possible departure from $u$ (to reach $v$) and the earliest possible arrival at $w$ (when coming from $v$):

\begin{itemize}
    \item The earliest time to reach node $w$ from $u$ equals that from $v$.
    \item The latest time one can leave $u$ to arrive at $w$ matches the latest departure time to reach $v$.
\end{itemize}

The TCC value for $v$ is the total number of pairs covered by $v$, divided by the total number of node pairs. This value reflects the proportion of node pairs for which $v$ is on at least one fastest temporal path. A high TCC value signifies redundancy in the network’s information flow, indicating multiple fastest routes exist for information transfer and that the temporal vertex frequently participates in these optimal paths.
\begin{table}[H]
\small
\centering
\caption{Summary of Centrality Measures for Temporal Networks}
\label{table:temporalcentrality}
\resizebox{\textwidth}{!}{
\begin{tabular}{p{3.5cm} p{3.5cm} p{4.5cm} p{4.5cm} p{4.5cm}}
\toprule
\textbf{Reference} & \textbf{Centrality Measure} & \textbf{Objective} & \textbf{Limitation} & \textbf{Application} \\
\midrule
Wang \textit{et al.} \cite{wang2017new} & Efficiency Centrality & To determine the most influential nodes and spreaders in complex networks by quantifying how their removal impacts the overall efficiency of the network & Computationally expensive & Viral marketing, disease spreading \\

Oettershagen \textit{et al.} \cite{oettershagen2022temporal} & Temporal Walk Centrality & To quantify a node's importance by assessing its capability to acquire and disseminate information within a temporal network & High computational complexity & Information spreading \\

Lerman \textit{et al.} \cite{lerman2010centrality} & Dynamic Centrality & To propose a dynamic centrality metric that incorporates temporal ordering and memory of interactions, improving over static path-based centrality in evolving networks. & Requires rich temporal data for accurate parameter estimation. & Citation networks, social and communication networks \\

Huang \textit{et al.} \cite{huang2017dynamic} & Dynamic-Sensitive Centrality & To combine temporal activity and structural position for evaluating time-varying node influence. & Parameter-dependent; requires diffusion-model validation. & Epidemic spreading, dynamic communication, and social networks \\

Takaguchi \textit{et al.} \cite{takaguchi2016coverage} & Temporal Coverage Centrality & To measure how often a temporal vertex appears on the fastest temporal paths between node pairs, indicating its role in efficient information transfer. & May overestimate importance as it includes redundant vertices that are not crucial for fastest communication. & Information or influence propagation, epidemic control \\

Takaguchi \textit{et al.} \cite{takaguchi2016coverage} & Temporal Boundary Coverage Centrality & To identify temporal vertices that are essential boundaries in fastest temporal paths, marking critical time points for communication. & May miss indirectly influential vertices and focuses only on strict boundary conditions. & Transportation and logistics networks, temporal bottleneck detection \\

Rocha \textit{et al.} \cite{rocha2014random} & TempoRank & To quantify node importance in dynamic networks using the stationary density of a random walk with periodic boundary conditions, thereby linking structural and temporal diffusion characteristics & Assumes periodic boundary conditions, which may not always reflect real non-repetitive temporal dynamics. & Human contact, communication, transportation, and epidemic networks \\
\bottomrule
\end{tabular}}
\end{table}

\subsection{Temporal Boundary Coverage Centrality (TBCC)}

Temporal Boundary Coverage Centrality (TBCC) extends Temporal Coverage Centrality (TCC) by introducing an additional boundary condition to identify temporal vertices that are essential for fastest temporal paths \cite{takaguchi2016coverage}. A vertex $v=(v,\tau)$ covers a node pair $(u,w)$ if it satisfies the TCC conditions and either the earliest arrival from $u$ to $v$ or the latest departure from $v$ to $w$ occurs exactly at time $\tau$. This stricter criterion filters out non-essential intermediate vertices and highlights critical temporal nodes whose removal would significantly delay information propagation.

\subsection{TempoRank (TR)}

TempoRank is a temporal centrality measure based on random walks in time-varying networks \cite{rocha2014random}. Unlike static random-walk centrality, it accounts for both network structure and temporal ordering of interactions. The temporal network is represented as a sequence of adjacency matrices $\{A^1, A^2, \dots, A^T\}$, where $A_{xy}(t)$ denotes interactions between nodes $x$ and $y$ at time $t$. For each snapshot, a transition probability matrix $B(t)$ is constructed using a sojourn probability $q$, which controls the probability of remaining at the current node. Multiplying these matrices chronologically yields the overall transition matrix:
\[
P_{\text{tp}} = B^{r}(t)B^{r-1}(t)\cdots B^{1}(t).
\]

The TempoRank centrality is obtained from the stationary distribution of the random walker, computed as the leading eigenvector of $P_{\text{tp}}$. For computational efficiency, an approximate solution can also be derived by aggregating temporal pathways across snapshots. By incorporating both temporal ordering and connectivity patterns, TempoRank effectively identifies structurally and dynamically important nodes in evolving networks.

\section{Real-World Applications and Use-Cases of Temporal Centrality Measures}
\label{sec:usecases}
Temporal centrality metrics have been widely utilised across a variety of real-world domains where interactions evolve over time. The choice of an appropriate centrality measure in temporal networks relies on several factors, including the nature of the dataset, the temporal resolution, the application goal, and the computational constraints. Since each temporal centrality captures different aspects of influence, flow, and persistence, understanding the context of analysis is essential for meaningful interpretation. This section provides a practical guide for selecting appropriate temporal centrality measures, temporal network models, and taxonomic considerations based on specific application scenarios. Table~\ref{tab:usecase} presents a summary of major application domains along with the temporal centrality measures that are most suitable for each setting.
\subsection{Healthcare and Contact Networks} Temporal centrality measures play a critical role in analyzing contact networks in healthcare environments. They are used to identify potential super-spreaders who may facilitate the rapid transmission of infections among patients and other healthcare workers. These insights can support the design of targeted intervention strategies, such as vaccination prioritization or contact isolation, to prevent epidemic outbreaks in healthcare settings. Temporal centrality measures that capture frequent interactions and temporal reachability, such as TWC, TCC and TSDC, are highly effective in identifying potential super-spreaders. Due to the rapidly evolving contact patterns, measures that rely on temporal paths outperform those based on static structure. This suggests that in epidemic modeling and healthcare settings, time-aware and connectivity-driven centrality measures are crucial for effective intervention strategies.
\subsection{Communication and Online Interaction Networks} Communication networks represent one of the most widely studied applications of temporal network analysis. In such networks, temporal centrality measures can be employed to determine influential users who play a significant role in disseminating information or facilitating communication. Centrality measures that account for temporal ordering and cumulative influence, such as TWC, TSDC and eigen-vector-based centrality are generally more effective in capturing the dynamic flow of information. In contrast, other temporal measures, such as temporal closeness or efficiency centrality, emphasise shortest paths and may miss recurring mediators.
\subsection{Financial and Economic Transaction Networks} Another important application domain for temporal centrality measures is financial and economic transaction networks, where interactions represent monetary transfers or trust relationships occurring over time. In such systems, temporal analysis can reveal influential participants who significantly affect transaction flows and trust propagation. Temporal centrality measures such as TWC, TSDC and betweenness, that capture sustained influence and temporal paths are better suited for identifying key actors in financial and transaction networks.
\begin{table}[H]
\small
\centering
\caption{Summary of Application Domains and Suitable Temporal Centrality Measures}
\label{tab:usecase}
\resizebox{\textwidth}{!}{
\begin{tabular}{p{3.5cm} p{3.5cm} p{4.5cm} p{3.5cm} p{4.5cm}}
\toprule
\textbf{Network Domain} & \textbf{Scenario} & \textbf{Recommended Centrality Measure} & \textbf{Graph Type} & \textbf{Reasoning} \\
\midrule
Healthcare and Contact Networks & Identify epidemic spreaders, model disease propagation, support epidemic control strategies & Time-scale Degree Centrality, Temporal Walk Centrality, Temporal Coverage Centrality & Undirected temporal graph &  These measures model temporal reachability and path-based influence \\
Communication and Online Interaction Networks & Identify evolving influencers or information mediators & Temporal Walk Centrality, TSDC, Supracentrality & Directed temporal graph & These measures capture temporal ordering and cumulative influence over time \\

Financial and Economic Transaction Networks & Identify trust hubs, detect influential traders, systemic risk contributors, and evolving market influence & TWC, Betweeness,  Efficiency Centrality, Temporal PageRank & Weighted directed temporal graph (edges represent time-stamped monetary or trade transactions) & Efficiency-based centralities assess network robustness against temporal shocks or cascading failures. \\



\bottomrule
\end{tabular}}
\end{table}

\par
These diverse application domains highlight the importance of temporal centrality measures in analyzing dynamic interaction networks. From identifying influential communicators in online platforms to detecting super-spreaders in contact networks and trust hubs in financial systems, temporal centrality provides valuable insights into how influence evolves over time. Motivated by these real-world applications, the experimental evaluation conducted in this study examines the effectiveness of several temporal centrality measures across representative datasets drawn from these domains.
\section{Experimental Evaluation} 
\label{sec:exper_eval}
In this section, we perform experimental evaluations to compare various temporal centrality measures discussed in this survey. The experiments aim to analyse their effectiveness in identifying influential nodes across diverse application-domains using real-world datasets and appropriate evaluation metrics.
\subsection{Setup} All experiments were executed on a standard computing environment with multi-core processor and 16 GB of RAM. The implementation was carrried out in Python 3.11. 
\subsubsection{Datasets}
To assess the effectiveness of the temporal centrality measures in a systematic manner, experiments were carried out using three real-world temporal network datasets discussed below. 
Table \ref{tab:topologyRW} summarizes the basic statistics of these networks.
\begin{table}[H]
\small
\centering
\caption{Statistical Overview of the three real-world networks.}
\label{tab:topologyRW}
\resizebox{\textwidth}{!}{
\begin{tabular}{p{3.5cm} p{3.5cm} p{1.5cm} p{1.5cm} p{1.5cm}p{1.5cm} p{1.5cm}}
\toprule
\textbf{Network} & \textbf{Application Domain} & \textbf{$n$} & \textbf{$m$} &\textbf{Snapshots} & \textbf{Time-window(in days)} & \textbf{Seed Nodes}\\
\midrule
Hospital ward & \makecell{Healthcare Contact\\ Networks} & 75 & 32424 & 10 & 0.3 & 4 \\
CollegeMsg & \makecell{Online Interaction\\ Networks} & 1899 & 59835 & 19 & 10 & 95\\
Bitcoin otc & \makecell{Financial Transaction\\ Networks} & 5881 & 35592 & 24 & 80 & 294 \\
\bottomrule
\end{tabular}}
\end{table}
\begin{enumerate}
\item Hospital Ward Dynamic Contact Network\footnote{https://sociopatterns.org/datasets/hospital-ward-dynamic-contact-network/}: It captures the time-varying contact interactions among 29 patients, interactions between patients and 46 healthcare workers, and contacts among the healthcare workers within a hospital ward located in France.
\item CollegeMsg Temporal Network \footnote{https://snap.stanford.edu/data/CollegeMsg.html}:It consists of private messages exchanged on an online social network at the University of California, Irvine. A temporal edge $(x,y,t)$ indicates that user $ x$ sent a message to user $y$ at time $t$.
\item Bitcoin Otc Trust Network \footnote{https://snap.stanford.edu/data/soc-sign-bitcoin-otc.html}: This dataset represents a who-trusts-whom network of Bitcoin traders on the Bitcoin OTC platform. 
\end{enumerate}

\subsubsection{Centrality Measures Compared}
We evaluate a set of representative temporal centrality metrics that were discussed in Section \ref{sec: CentralityM} and \ref{sec: Centrality Temporal}. These metrics capture different aspects of node significance in temporal networks and are widely used to find important nodes in dynamic systems. A brief description of the compared centrality measures is provided below.
\begin{enumerate}
    \item Time-Scale Degree Centraliy (TSDC) \cite{uddin2011time}: This measure extends traditional degree centrality to temporal networks by incorporating both the presence of links and the duration of interactions between actors .
    \item Temporal Coreness Centrality \cite{li2013efficient}: An efficient algorithm is proposed for maintaining the core numbers of nodes in a temporal network. The main observation is that updates caused by edge insertions or deletions affect only a limited set of nodes. Accordingly, the algorithm identifies these nodes and efficiently recalculates their core numbers.
    \item CENDY \cite{yen2013efficient}: It provides an efficient mechanism for updating both closeness centrality and average path length in dynamic networks under edge insertion and deletion operations .
    \item QUBE \cite{lee2012qube}: It minimises the search space by identifying a candidate set of vertices whose betweenness centralities need to be updated. The betweenness values are then recomputed by considering only these candidate vertices .
    \item Supracentrality \cite{taylor2017eigenvector}: It is a generalized framework for eigenvector-based centrality measures in temporal networks. In this approach, a dynamic network with $N$ nodes is represented as a sequence of $T$ layers corresponding to different time intervals. The centrality matrices of these layers are combined into a supracentrality matrix of size $NT \times NT$, whose dominant eigenvector determines the centrality of each node at every time step.
    \item Incremental PageRank (Incremental PRank) \cite{desikan2005incremental}: It is an incremental approach for computing PageRank in dynamic networks. This algorithm recomputes PageRank only for the portion of the graph that has changed, thereby reducing computational cost.
    \item Efficiency Centrality (EffC) \cite{wang2017new}: It is a global centrality metric which identifies influential nodes based on their contribution to the overall efficiency of a network. It quantifies the extent to which network efficiency decreases when a node is removed, indicating the node's significance in maintaining efficient communication.
    \item Temporal Walk Centrality (TWC)\cite{oettershagen2022temporal}: It measures the influence of nodes in evolving networks by counting the number of time-respecting walks that originate from each node.
    \item Temporal Coverage Centrality (TCC) \cite{takaguchi2016coverage}: It quantifies the significance of nodes in temporal graphs by measuring how often a node appears on the earliest time-respecting paths between other node pairs.
\end{enumerate}
\subsubsection{Evaluation Metrics} To evaluate the effectiveness of the compared temporal centrality measures, we employ two metrics: Mean Fraction of Infected (MFI) and Kendall’s $\tau$ correlation coefficient.
\begin{enumerate}
    \item Mean Fraction of Infected (MFI): This metric is used to determine the spreading capability of nodes identified by different centrality metrics.  For each method, the top-$k$ ranked nodes are selected as initial seeds and a spreading process is simulated on the temporal network. The number of infected nodes is recorded at each time step and averaged over multiple simulation runs.
    \item Kendall’s $\tau$ Correlation: It is used to measure the similarity between the node rankings produced by different centrality measures. Higher Kendall’s $\tau$ values indicate stronger agreement between node rankings.
    \item Runtime Analysis: To evaluate scalability, the execution time of each centrality measure was recorded for every temporal snapshot under the same hardware and software environment. Runtime was measured in seconds and reported on a logarithmic scale to facilitate comparison across methods with substantially different computational requirements. This analysis provides insight into the practical feasibility of applying temporal centrality measures to evolving networks of varying size and complexity.
   
\end{enumerate}
\subsection{Influence Spreading Experiment} To evaluate the capability of temporal centrality measures in identifying influential nodes, influence spreading experiments are performed using the SI and SIS diffusion models. For each method, the top 5\% ranked nodes are selected as initial seeds. Infection transmission probability is set to $\beta=0.01$, while the SIS recovery probability is fixed at $\gamma=0.02$. Each experiment is repeated over 1000 Monte Carlo simulations, and the Mean Fraction of Infected (MFI) is reported over time.
\subsection{Results and Discussion} In this subsection, we present and analyze the results obtained from the influence spreading experiments, ranking correlation analysis and runtime complexity evaluation. The performance of the compared centrality metrics is evaluated using the mean fraction of infected (MFI) curves under the SI and SIS models, Kendall’s $\tau$ correlation coefficient to examine the similarity between the generated node rankings, and runtime measurements to evaluate the computational efficiency and scalability of the methods across different temporal network datasets.
\subsubsection{Influence Spreading Results}
\begin{figure}[H]
    \centering
     \caption{MFI curves under the SI model across temporal datasets.}
\label{fig:MFI_SI}
    \begin{subfigure}[t]{0.32\linewidth}
    \centering
    \includegraphics[width=0.85\linewidth]{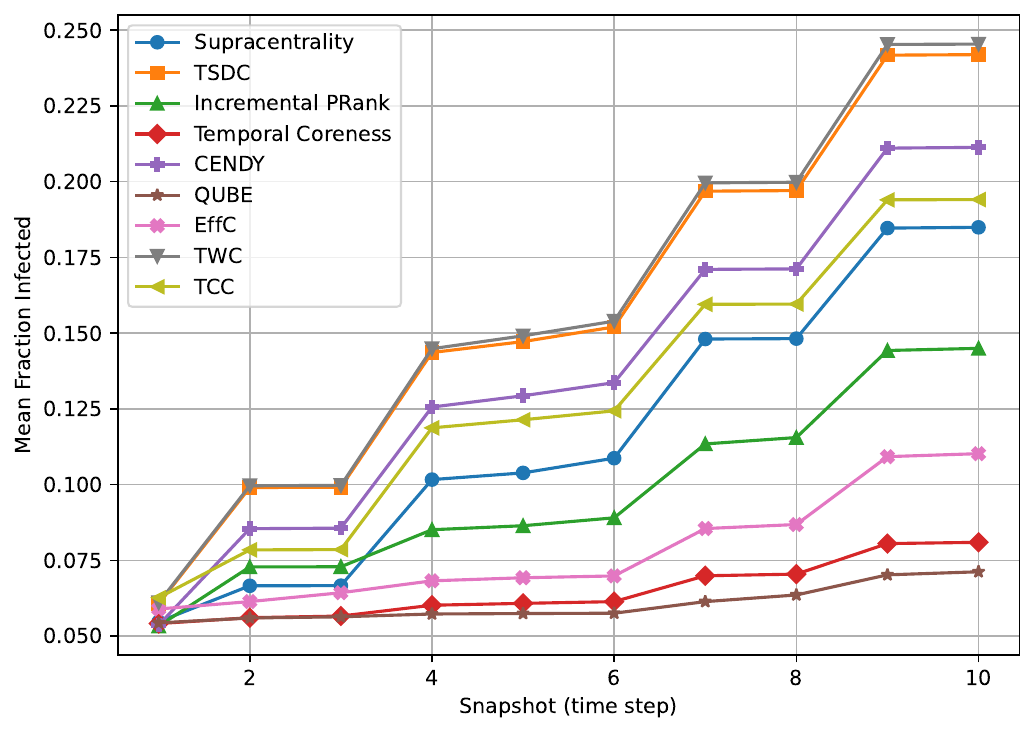}
    \caption{Hospital Ward Dataset}
    \label{fig:HospSI}
    \end{subfigure}
    \hfill
    \begin{subfigure}[t]{0.32\linewidth}
    \centering
    \includegraphics[width=0.85\linewidth]{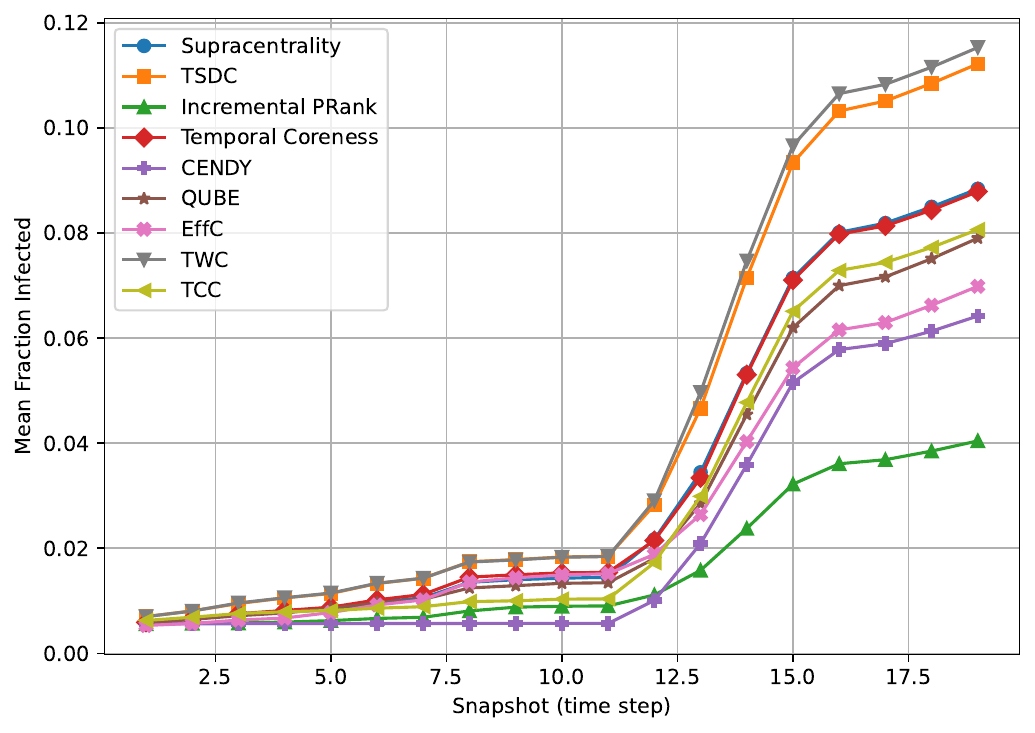}
    \caption{CollegeMsg Dataset}
    \label{fig:CollegeMsgSI}
    \end{subfigure}
    \hfill
     \begin{subfigure}[t]{0.32\linewidth}
    \centering
    \includegraphics[width=0.85\linewidth]{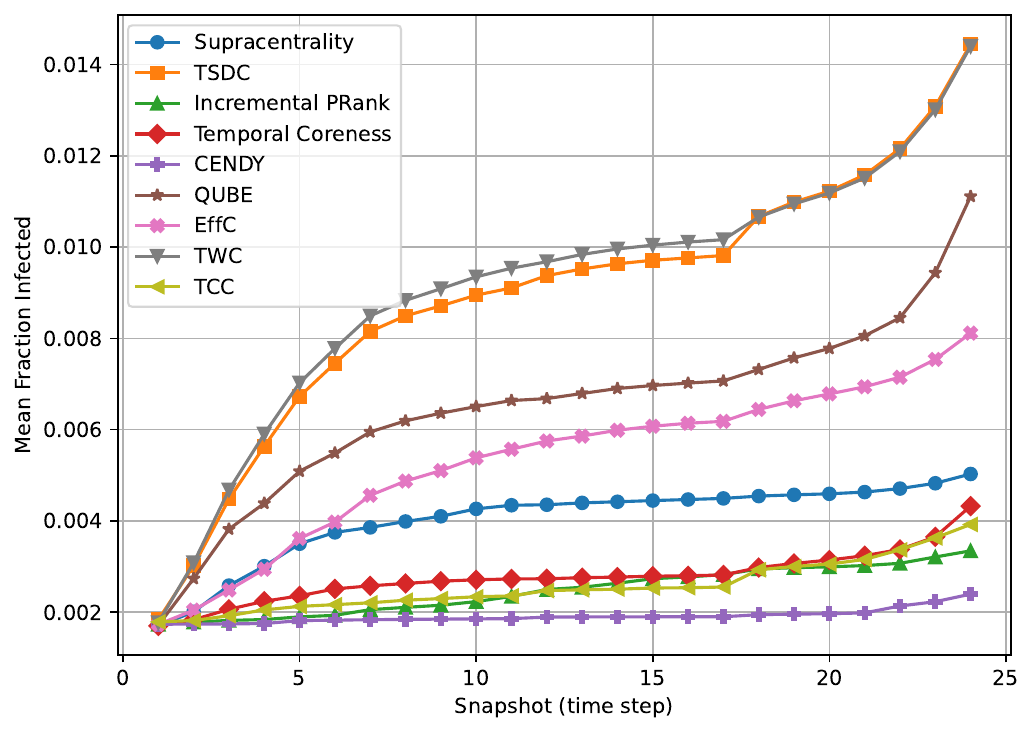}
    \caption{BitcoinOtc Dataset}
    \label{fig:BitcoinSI}
    \end{subfigure}
\end{figure}

\begin{figure}[H]
    \centering
    \caption{MFI curves under the SIS model across temporal datasets.}
    \label{fig:MFI_SIS} 
    \begin{subfigure}[t]{0.32\linewidth}
    \centering
     \includegraphics[width=0.85\linewidth]{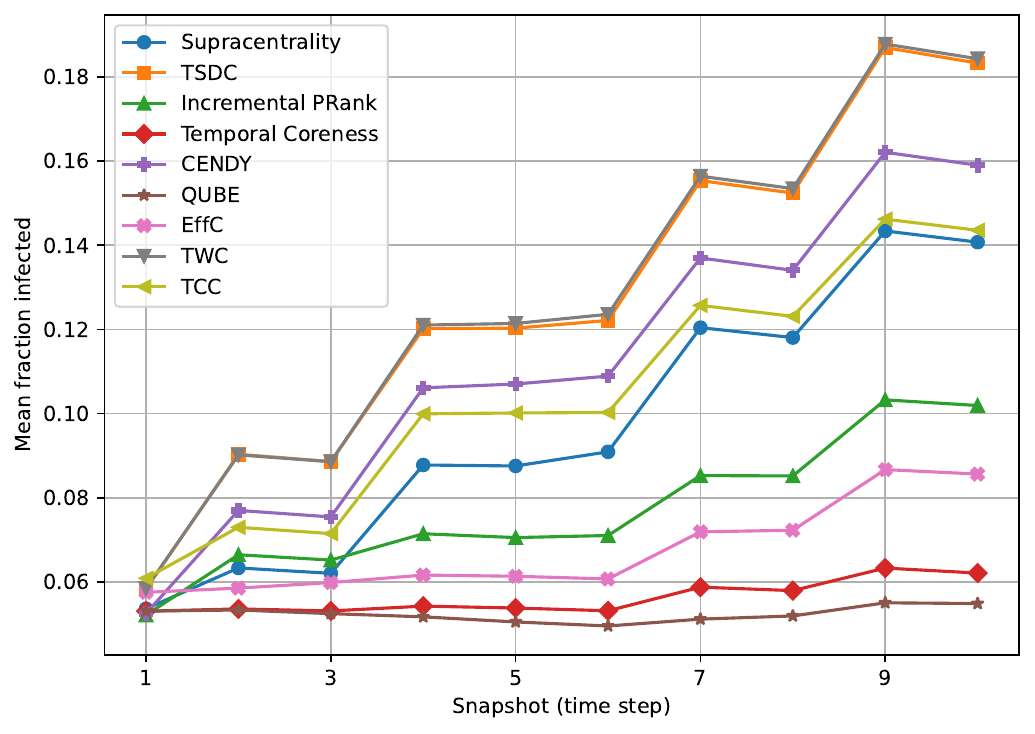}
    \caption{Hospital Ward Dataset}
    \label{fig:HospitalSIS}
    \end{subfigure}
    \hfill
     \begin{subfigure}[t]{0.32\linewidth}
    \centering
      \includegraphics[width=0.85\linewidth]{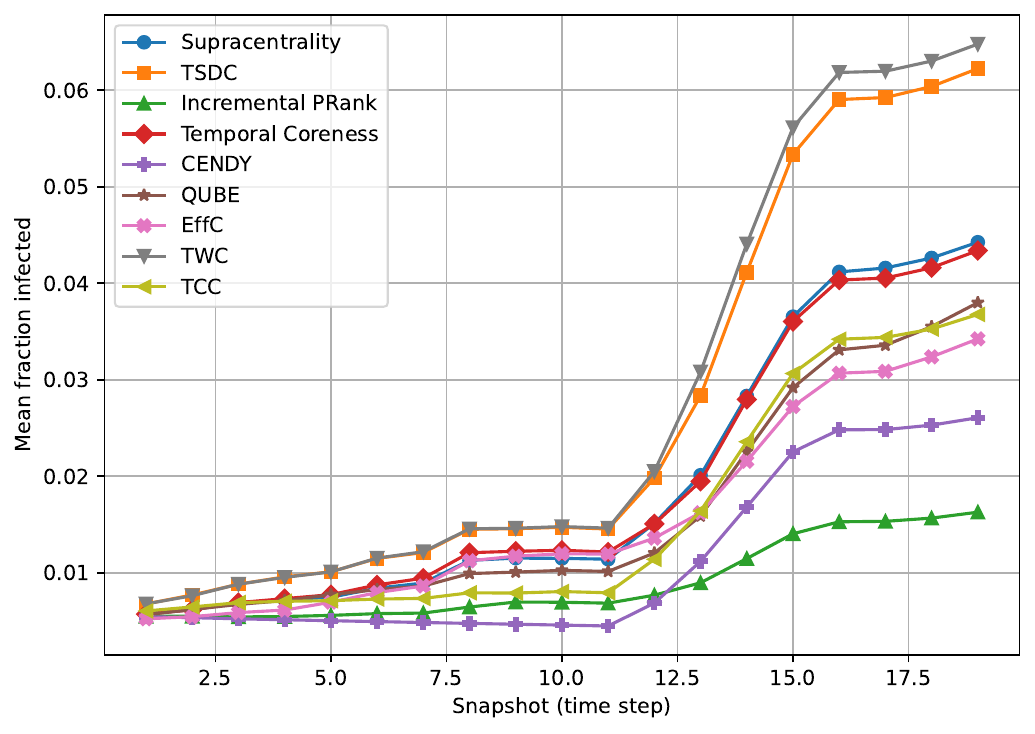}
    \caption{CollegeMsg Dataset}
    \label{fig:CollegeMsgSIS}
    \end{subfigure}
    \hfill
     \begin{subfigure}[t]{0.32\linewidth}
    \centering
    \includegraphics[width=0.85\linewidth]{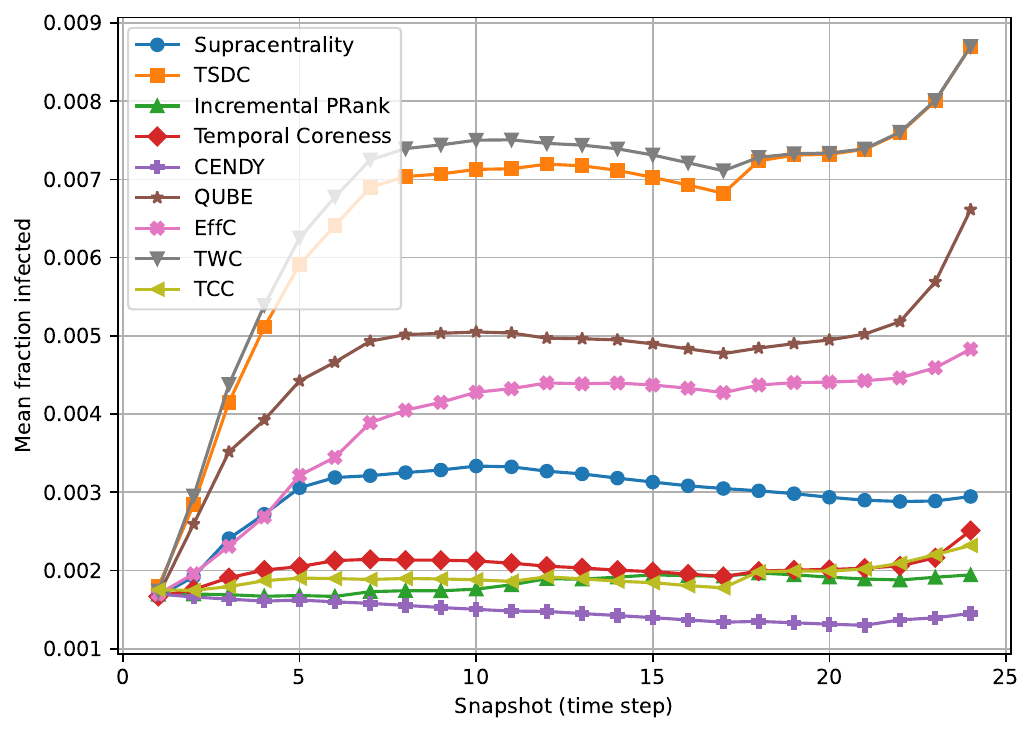}
    \caption{BitcoinOtc Dataset}
    \label{fig:BitcoinOtcSIS}
    \end{subfigure}
  \end{figure}

The influence spreading results highlight that diffusion dynamics are strongly shaped by the underlying network structure and interaction patterns. Dense contact networks, such as the Hospital dataset, exhibit rapid and large-scale spreading, whereas online interaction networks (CollegeMsg) show delayed growth with clear phase transitions. In contrast, the BitcoinOtc network demonstrates smoother and slower diffusion, reflecting its trust-based interaction dynamics. Under the SI model, the diffusion behaviour for the Hospital Ward, CollegeMsg, and BitcoinOtc datasets is illustrated in Figure \ref{fig:HospSI}, Figure \ref{fig:CollegeMsgSI}, and Figure \ref{fig:BitcoinSI}, respectively. Similarly, the SIS-based diffusion results for the Hospital Ward, CollegeMsg, and BitcoinOtc datasets are presented in Figure \ref{fig:HospitalSIS}, Figure \ref{fig:CollegeMsgSIS}, and Figure \ref{fig:BitcoinOtcSIS}, respectively.\par

As observed in Figure \ref{fig:MFI_SI}(a-c) and Figure \ref{fig:MFI_SIS}(a–c), TWC and TSDC consistently achieve the highest diffusion performance under both SI and SIS models.. Their ability to capture temporal reachability and dynamic interaction patterns enables more effective identification of influential nodes. In contrast, methods such as Supracentrality, Incremental PageRank, and CENDY exhibit comparatively weaker spreading capability. The results further demonstrate that diffusion effectiveness strongly depends on the temporal characteristics and structural properties of the underlying network.
Overall, the results demonstrate that the effectiveness of temporal centrality measures is application-dependent, and approaches that incorporate temporal walk-based and time-aware mechanisms are more suitable for influence analysis in dynamic networks.
\subsubsection{Ranking Correlation Analysis} 
\begin{figure}[H]
    \centering
    \caption{Kendall’s $\tau$ correlation heatmaps across datasets.}
\label{fig:Kendall_Correlation}  
    \begin{subfigure}[t]{0.32\linewidth}
    \centering
    \includegraphics[width=\linewidth]{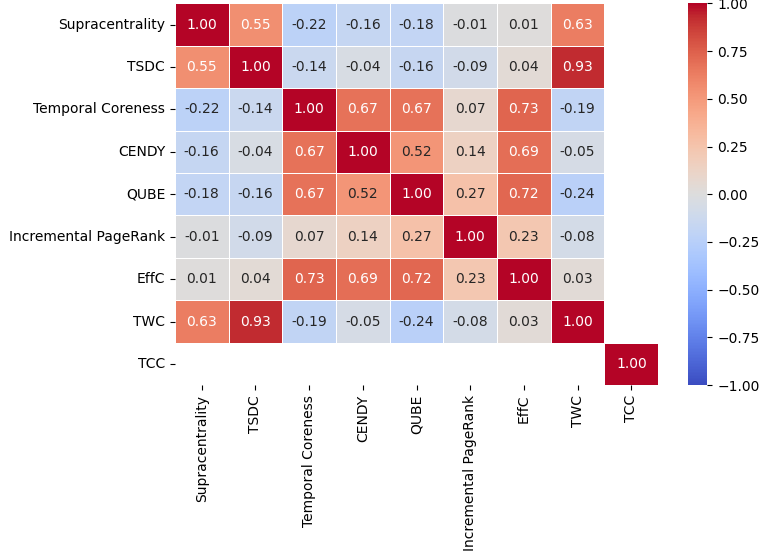}
    \caption{Hospital Ward Dataset}
    \label{fig:HospKendall}
    \end{subfigure}
    \hfill
      \begin{subfigure}[t]{0.32\linewidth}
    \centering
    \includegraphics[width=\linewidth]{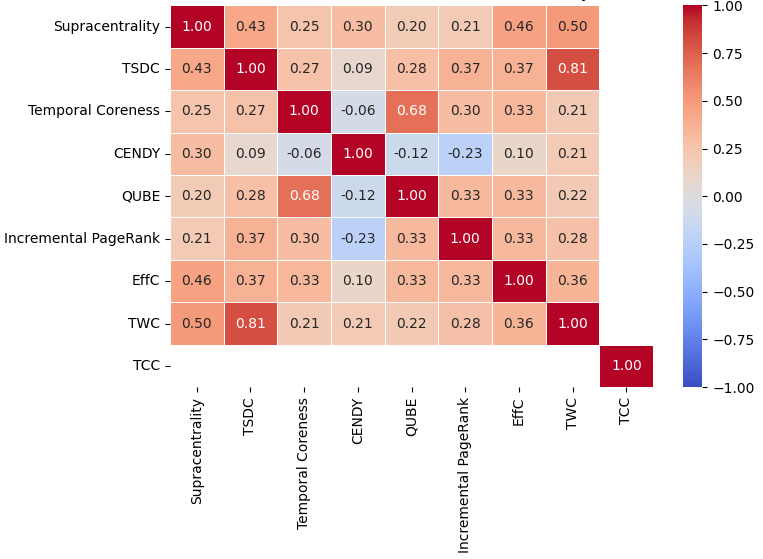}
    \caption{CollegeMsg Dataset}
    \label{fig:CollegeMsgKendall}
    \end{subfigure}
    \hfill
     \begin{subfigure}[t]{0.32\linewidth}
    \centering
    \includegraphics[width=\linewidth]{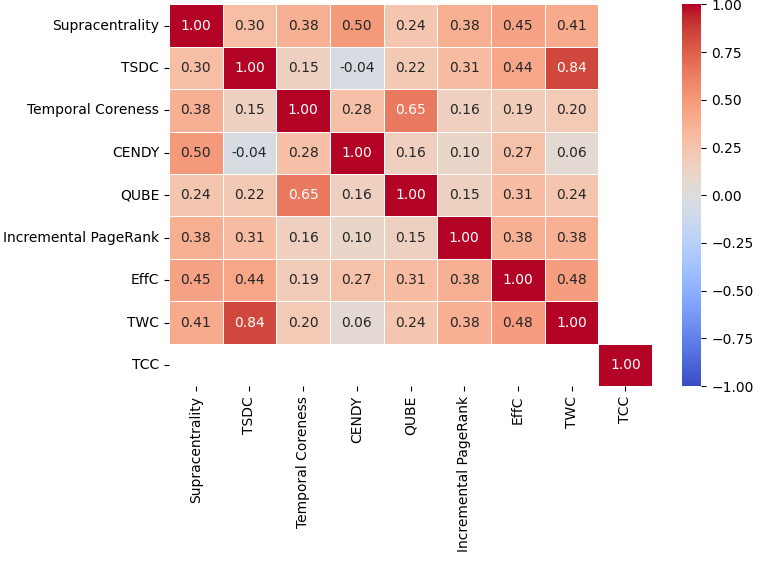}
    \caption{BitcoinOtc Dataset}
    \label{fig:BitcoinOtcKendall}
    \end{subfigure}
  \end{figure}

Kendall’s $\tau$ analysis complements diffusion-based evaluation by revealing whether high-performing centrality measures agree on node importance or capture fundamentally different influence patterns. Figure \ref{fig:HospKendall}, Figure \ref{fig:CollegeMsgKendall} and Figure \ref{fig:BitcoinOtcKendall} show the Kendall’s $\tau$ correlation heatmaps for the Hospital Ward, CollegeMsg, and BitcoinOtc datasets, respectively.\par
As shown in Figure \ref{fig:HospKendall}, a consistently strong correlation is observed between TSDC and TWC indicating that these methods tend to identify similar influential nodes in dense networks. This strong agreement indicates that both metrics rely on related temporal diffusion or connectivity patterns when identifying influential nodes. Additionally, Temporal Coreness and QUBE frequently exhibit moderate to strong correlations, implying that these methods capture comparable structural properties related to temporal connectivity and node influence. 
 
In contrast, CENDY often shows weak or even negative correlations with several other measures, indicating that it ranks nodes differently and captures alternative characteristics of influence within the network. Similarly, Incremental PageRank tends to exhibit moderate correlations with other measures, suggesting partial agreement but also highlighting differences in how influence is evaluated. Overall, Kendall’s $\tau$ analysis demonstrates that no single centrality measure fully aligns with all others, emphasizing that different measures reflect different dimensions of node significance in temporal networks.

\subsubsection{Runtime Analysis Results}

\begin{figure}[H]
    \centering
    \caption{Runtime Complexity across datasets.}
\label{fig:runtime_complexity}  
    \begin{subfigure}[t]{0.32\linewidth}
    \centering
    \includegraphics[width=\linewidth]{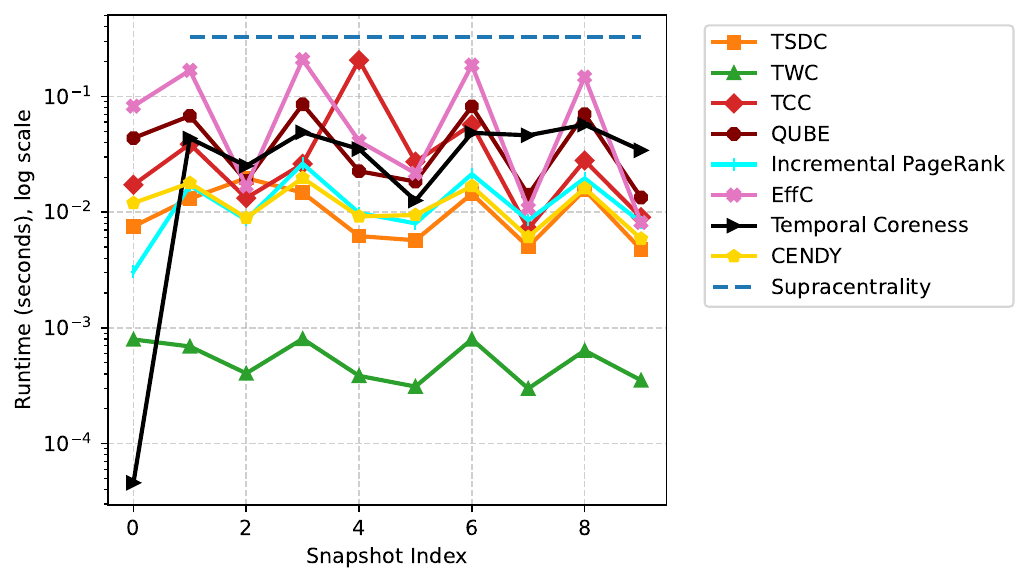}
    \caption{Hospital Ward Dataset}
    \label{fig:Hospruntime}
    \end{subfigure}
    \hfill
      \begin{subfigure}[t]{0.32\linewidth}
    \centering
    \includegraphics[width=\linewidth]{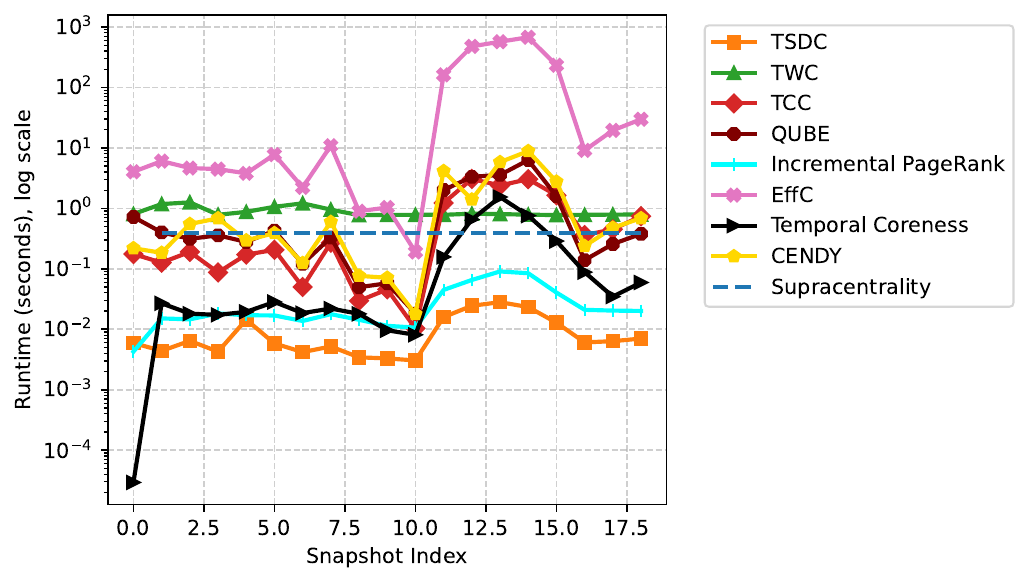}
    \caption{CollegeMsg Dataset}
    \label{fig:CollegeMsgruntime}
    \end{subfigure}
    \hfill
     \begin{subfigure}[t]{0.32\linewidth}
    \centering
    \includegraphics[width=\linewidth]{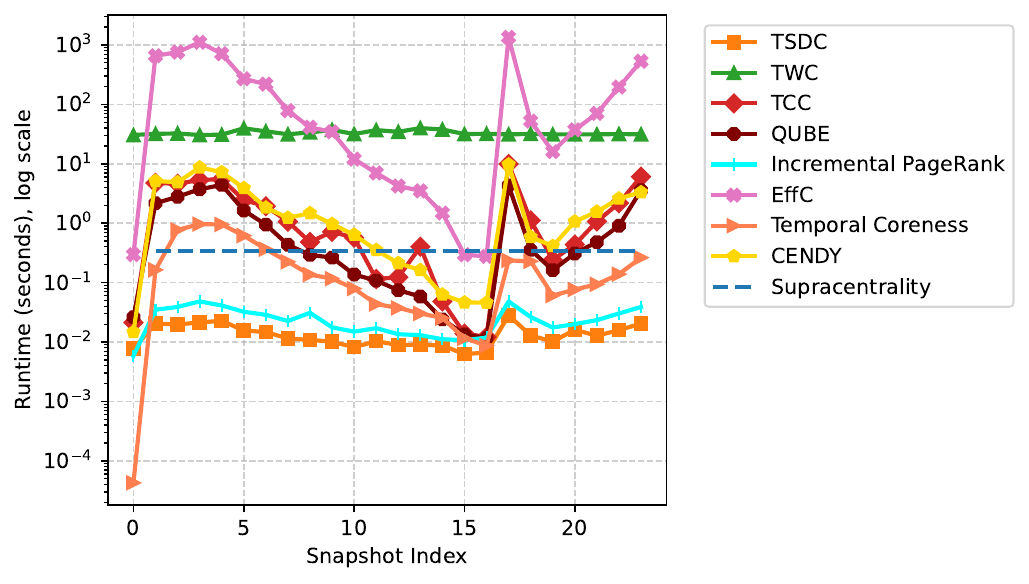}
    \caption{BitcoinOtc Dataset}
    \label{fig:BitcoinOtcruntime}
    \end{subfigure}
  \end{figure}

Figure \ref{fig:Hospruntime}, Figure \ref{fig:CollegeMsgruntime} , and Figure \ref{fig:BitcoinOtcruntime} present the runtime comparison of the evaluated temporal centrality measures on the Hospital Ward, CollegeMsg, and BitcoinOtc datasets, respectively. Overall, the results reveal substantial differences in computational cost among the compared methods, reflecting the complexity of their underlying mechanisms.\par
Across all datasets, TSDC consistently exhibits the lowest execution time, requiring only a fraction of a millisecond per snapshot. The stable runtime behaviour of TSDC indicates excellent scalability and makes it particularly suitable for large temporal networks and real-time applications. TWC and Incremental Pagerank also demonstrate low computational overhead, maintaining execution times in the order of milliseconds throughout the temporal evolution of the networks.\par
In contrast, Efficiency Centrality(EffC) and Supracentrality are among the most computationally demanding methods. EffC consistently records the highest runtime across most snapshots, while Supracentrality maintains a relatively high and stable execution cost due to the construction and processing of large temporal coupling structures. EffC exhibits noticeable fluctuations across snapshots, which can be attributed to variations in network structure and the repeated computation of network efficiency following node removals. These observations suggest that spectral-based approaches provide richer global information at the expense of increased computational requirements. Intermediate runtime behaviour is observed for QUBE, CENDY, Temporal Coreness, and Temporal Coverage Centrality (TCC). \par
A comparison across datasets further indicates that runtime generally increases with network size and temporal complexity. The BitcoinOtc and CollegeMsg datasets, which contain significantly larger numbers of nodes and interactions than the Hospital Ward dataset, exhibit greater computational demands for most methods. Nevertheless, the relative ordering of the centrality measures remains largely consistent across all datasets, demonstrating that their computational characteristics are primarily determined by the underlying algorithmic design rather than the specific network domain.
 These results highlight a clear trade-off between computational efficiency and methodological complexity. Local interaction-based approaches such as TSDC provide highly efficient execution and are well suited for large-scale temporal networks, whereas spectral and global-path-based methods offer richer structural information but require substantially greater computational resources. Therefore, the selection of an appropriate temporal centrality measure should consider not only influence identification accuracy but also the computational constraints of the target application.
\section{Open Challenges and Future Directions}
\label{sec: oc}
While significant efforts have been made in the development of temporal centrality measures, several critical challenges remain open for future exploration. Addressing these issues is essential to advance the applicability and scalability of centrality models in complex time-evolving systems. Beyond summarizing existing limitations, this section highlights forward-looking research directions that can guide the next generation of temporal centrality models. Table \ref{tab:open_challenges} summarizes the major open challenges and promising future research directions in temporal centrality analysis.

\subsection{Scalability and Computational Efficiency}
One of the foremost challenges in temporal centrality research is ensuring scalability as network size increases. Most current methods, such as betweenness \cite{tang2010analysing}, efficiency centrality \cite{wang2017new}, or supracentrality \cite{taylor2017eigenvector}\cite{yin2018inter} require repeated computations over multiple time layers or large supra-adjacency matrices, leading to exponential growth in memory and time complexity. Future research should focus on scalable approximation and incremental update methods for large-scale temporal networks.
\subsection{Real-Time and Streaming Centrality Computation}
Modern applications, such as epidemic monitoring, traffic control, and cybersecurity, require real-time detection of influential nodes as the network evolves. However, most temporal centrality measures are designed for retrospective analysis and cannot be directly applied in streaming environments, where edges arrive continuously over time rather than being analysed as a complete, static dataset. Developing adaptive streaming centrality algorithms capable of real-time updates remains an important research direction.
\subsection{Incorporating Contextual and Semantic Information}
A major limitation of existing temporal centrality measures is their reliance solely on topological and temporal features, ignoring contextual factors such as node attributes, interaction semantics, or external influences. Real-world systems often exhibit context-dependent behaviors; for example, influence in a communication network depends not only on message timing but also on content and trust. Future designs should integrate semantic embeddings or heterogeneous network attributes into temporal centrality frameworks, leading to context-aware influence quantification. Combining network structure with domain knowledge can yield more realistic and interpretable centrality outcomes.

\begin{table}[H]
\footnotesize
\centering
\caption{Open challenges and future research directions in temporal centrality analysis}
\label{tab:open_challenges}
\resizebox{\textwidth}{!}{
\begin{tabular}{p{5cm} p{6.5cm} p{7cm}}
\toprule
\textbf{Challenge} & \textbf{Description} & \textbf{Future Research Directions} \\
\midrule
Scalability and Computational Efficiency &
Temporal centrality computation becomes increasingly expensive as the number of nodes and temporal layers grows, often requiring repeated matrix operations or recomputation. &
Development of approximation methods, sampling strategies, low-rank updates, and parallel or distributed frameworks for large-scale temporal graphs. \\

Real-time and Streaming Computation &
Most existing temporal centrality measures operate offline and are not designed for continuously evolving edge streams or event-based data. &
Design of incremental and adaptive algorithms that update centrality scores online using streaming and event-driven computation models. \\

Contextual and Semantic Information &
Existing approaches rely mainly on temporal topology, ignoring node attributes, interaction semantics, and external contextual information. &
Integration of semantic embeddings, heterogeneous node attributes, and domain-specific information into temporal centrality frameworks. \\


\bottomrule
\end{tabular}}
\end{table}

\section{Conclusion}
\label{sec: conclusion}
The study of centrality in temporal networks has evolved from static graph-based metrics toward dynamic and time-aware formulations that more faithfully represent real-world interaction processes. This survey introduced a functional taxonomy that organizes existing temporal centrality measures according to their primary source of influence, providing a unified framework for understanding the relationships among diverse approaches and facilitating informed method selection. In addition, this survey reviewed both temporal adaptations of classical centrality measures, such as betweenness, coreness, degree, closeness, and PageRank, as well as specialized temporal metrics, including eigenvector-based, dynamic-sensitive, and efficiency-oriented approaches. Collectively, these methods demonstrate the critical role of temporal order, interaction duration, and causality in accurately assessing node importance in time-evolving networks. Despite their advantages, temporal extensions of traditional centrality measures often face practical challenges, including high computational complexity, sensitivity to temporal granularity, and limited scalability in large or streaming networks. More recent hybrid and dynamic approaches, particularly those grounded in spectral or efficiency-based reasoning, offer improved trade-offs between accuracy, adaptability, and computational feasibility, making them increasingly attractive for real-world applications.\par
Overall, the field is moving toward integrative frameworks that combine structural, temporal, and functional perspectives to better capture influence dynamics in evolving systems. Promising future directions include the development of scalable and real-time algorithms for temporal centrality maintenance, the incorporation of machine learning and temporal embeddings for predictive influence modeling, and the design of context-aware hybrid centrality measures that integrate structural, temporal, and semantic information. In addition, the establishment of standardized benchmark datasets and reproducible evaluation frameworks remains essential for enabling consistent comparison, validation, and informed method selection in temporal centrality research.
\bibliographystyle{unsrt}

\bibliography{cas-refs}

@article{holme2012temporal,
  title={Temporal networks},
  author={Holme, Petter and Saram{\"a}ki, Jari},
  journal={Physics reports},
  volume={519},
  number={3},
  pages={97--125},
  year={2012},
  publisher={Elsevier}
}

@article{saxena2020centrality,
  title={Centrality measures in complex networks: A survey},
  author={Saxena, Akrati and Iyengar, Sudarshan},
  journal={arXiv preprint arXiv:2011.07190},
  year={2020}
}

@inproceedings{uddin2011time,
  title={Time scale degree centrality: A time-variant approach to degree centrality measures},
  author={Uddin, Shahadat and Hossain, Liaquat},
  booktitle={2011 International Conference on Advances in Social Networks Analysis and Mining},
  pages={520--524},
  year={2011},
  organization={IEEE}
}

@article{lewis2008tastes,
  title={Tastes, ties, and time: A new social network dataset using Facebook. com},
  author={Lewis, Kevin and Kaufman, Jason and Gonzalez, Marco and Wimmer, Andreas and Christakis, Nicholas},
  journal={Social networks},
  volume={30},
  number={4},
  pages={330--342},
  year={2008},
  publisher={Elsevier}
}

@article{borgatti2009network,
  title={Network analysis in the social sciences},
  author={Borgatti, Stephen P and Mehra, Ajay and Brass, Daniel J and Labianca, Giuseppe},
  journal={science},
  volume={323},
  number={5916},
  pages={892--895},
  year={2009},
  publisher={American Association for the Advancement of Science}
}

@article{uddin2014new,
  title={New direction in degree centrality measure: Towards a time-variant approach},
  author={Uddin, Shahadat and Hossain, Liaquat and Wigand, Rolf T},
  journal={International Journal of Information Technology \& Decision Making},
  volume={13},
  number={04},
  pages={865--878},
  year={2014},
  publisher={World Scientific}
}

@inproceedings{ferreira2011evaluating,
  title={Evaluating external researchers impact on a research community social capital with social network analysis},
  author={Ferreira, Ana Sofia and Fernandes, Carlos and de Castro Neto, Miguel},
  booktitle={6th Iberian Conference on Information Systems and Technologies (CISTI 2011)},
  pages={1--5},
  year={2011},
  organization={IEEE}
}

@inproceedings{lerman2010centrality,
  title={Centrality metric for dynamic networks},
  author={Lerman, Kristina and Ghosh, Rumi and Kang, Jeon Hyung},
  booktitle={Proceedings of the Eighth Workshop on Mining and Learning with Graphs},
  pages={70--77},
  year={2010}
}

@article{bullmore2009complex,
  title={Complex brain networks: graph theoretical analysis of structural and functional systems},
  author={Bullmore, Ed and Sporns, Olaf},
  journal={Nature reviews neuroscience},
  volume={10},
  number={3},
  pages={186--198},
  year={2009},
  publisher={Nature Publishing Group UK London}
}

@article{jordan2008identifying,
  title={Identifying important species: linking structure and function in ecological networks},
  author={Jord{\'a}n, Ferenc and Okey, Thomas A and Bauer, Barbara and Libralato, Simone},
  journal={Ecological Modelling},
  volume={216},
  number={1},
  pages={75--80},
  year={2008},
  publisher={Elsevier}
}

@article{freeman2002centrality,
  title={Centrality in social networks: Conceptual clarification},
  author={Freeman, Linton C and others},
  journal={Social network: critical concepts in sociology. Londres: Routledge},
  volume={1},
  number={3},
  pages={238--263},
  year={2002}
}

@inproceedings{yen2013efficient,
  title={An efficient approach to updating closeness centrality and average path length in dynamic networks},
  author={Yen, Chia-Chen and Yeh, Mi-Yen and Chen, Ming-Syan},
  booktitle={2013 IEEE 13th International Conference on Data Mining},
  pages={867--876},
  year={2013},
  organization={IEEE}
}

@article{sariyuce2013incremental,
  title={Incremental algorithms for network management and analysis based on closeness centrality},
  author={Sariyuce, Ahmet Erdem and Kaya, Kamer and Saule, Erik and Catalyurek, Umit V},
  journal={arXiv preprint arXiv:1303.0422},
  year={2013}
}

@inproceedings{tang2009temporal,
  title={Temporal distance metrics for social network analysis},
  author={Tang, John and Musolesi, Mirco and Mascolo, Cecilia and Latora, Vito},
  booktitle={Proceedings of the 2nd ACM workshop on Online social networks},
  pages={31--36},
  year={2009}
}

@inproceedings{tang2010analysing,
  title={Analysing information flows and key mediators through temporal centrality metrics},
  author={Tang, John and Musolesi, Mirco and Mascolo, Cecilia and Latora, Vito and Nicosia, Vincenzo},
  booktitle={Proceedings of the 3rd workshop on social network systems},
  pages={1--6},
  year={2010}
}

@article{bavelas1950communication,
  title={Communication patterns in task-oriented groups.},
  author={Bavelas, Alex},
  journal={Journal of the acoustical society of America},
  year={1950},
  publisher={Acoustical Society of American}
}

@article{latora2001efficient,
  title={Efficient behavior of small-world networks},
  author={Latora, Vito and Marchiori, Massimo},
  journal={Physical review letters},
  volume={87},
  number={19},
  pages={198701},
  year={2001},
  publisher={APS}
}

@article{wang2017new,
  title={A new measure of identifying influential nodes: Efficiency centrality},
  author={Wang, Shasha and Du, Yuxian and Deng, Yong},
  journal={Communications in Nonlinear Science and Numerical Simulation},
  volume={47},
  pages={151--163},
  year={2017},
  publisher={Elsevier}
}

@article{freeman1977set,
  title={A set of measures of centrality based on betweenness},
  author={Freeman, Linton C},
  journal={Sociometry},
  pages={35--41},
  year={1977},
  publisher={JSTOR}
}

@article{newman2005measure,
  title={A measure of betweenness centrality based on random walks},
  author={Newman, Mark EJ},
  journal={Social networks},
  volume={27},
  number={1},
  pages={39--54},
  year={2005},
  publisher={Elsevier}
}

@inproceedings{kas2013incremental,
  title={Incremental algorithm for updating betweenness centrality in dynamically growing networks},
  author={Kas, Miray and Wachs, Matthew and Carley, Kathleen M and Carley, L Richard},
  booktitle={Proceedings of the 2013 IEEE/ACM international conference on advances in social networks analysis and mining},
  pages={33--40},
  year={2013}
}

@inproceedings{green2012fast,
  title={A fast algorithm for streaming betweenness centrality},
  author={Green, Oded and McColl, Robert and Bader, David A},
  booktitle={2012 International Conference on Privacy, Security, Risk and Trust and 2012 International Confernece on Social Computing},
  pages={11--20},
  year={2012},
  organization={IEEE}
}

@article{kourtellis2015scalable,
  title={Scalable online betweenness centrality in evolving graphs},
  author={Kourtellis, Nicolas and Morales, Gianmarco De Francisci and Bonchi, Francesco},
  journal={IEEE Transactions on Knowledge and Data Engineering},
  volume={27},
  number={9},
  pages={2494--2506},
  year={2015},
  publisher={IEEE}
}

@inproceedings{xing2004weighted,
  title={Weighted pagerank algorithm},
  author={Xing, Wenpu and Ghorbani, Ali},
  booktitle={Proceedings. Second Annual Conference on Communication Networks and Services Research, 2004.},
  pages={305--314},
  year={2004},
  organization={IEEE}
}

@techreport{page1999pagerank,
  title={The PageRank citation ranking: Bringing order to the web.},
  author={Page, Lawrence and Brin, Sergey and Motwani, Rajeev and Winograd, Terry},
  year={1999},
  institution={Stanford infolab}
}

@article{lv2019pagerank,
  title={PageRank centrality for temporal networks},
  author={Lv, Laishui and Zhang, Kun and Zhang, Ting and Bardou, Dalal and Zhang, Jiahui and Cai, Ying},
  journal={Physics Letters A},
  volume={383},
  number={12},
  pages={1215--1222},
  year={2019},
  publisher={Elsevier}
}

@inproceedings{oettershagen2022temporal,
  title={Temporal walk centrality: ranking nodes in evolving networks},
  author={Oettershagen, Lutz and Mutzel, Petra and Kriege, Nils M},
  booktitle={Proceedings of the ACM Web conference 2022},
  pages={1640--1650},
  year={2022}
}

@inproceedings{desikan2005incremental,
  title={Incremental page rank computation on evolving graphs},
  author={Desikan, Prasanna and Pathak, Nishith and Srivastava, Jaideep and Kumar, Vipin},
  booktitle={Special interest tracks and posters of the 14th International Conference on World Wide Web},
  pages={1094--1095},
  year={2005}
}

@article{seidman1983network,
  title={Network structure and minimum degree},
  author={Seidman, Stephen B},
  journal={Social networks},
  volume={5},
  number={3},
  pages={269--287},
  year={1983},
  publisher={Elsevier}
}

@article{katz1953new,
  title={A new status index derived from sociometric analysis},
  author={Katz, Leo},
  journal={Psychometrika},
  volume={18},
  number={1},
  pages={39--43},
  year={1953},
  publisher={Springer-Verlag}
}

@article{taylor2017eigenvector,
  title={Eigenvector-based centrality measures for temporal networks},
  author={Taylor, Dane and Myers, Sean A and Clauset, Aaron and Porter, Mason A and Mucha, Peter J},
  journal={Multiscale Modeling \& Simulation},
  volume={15},
  number={1},
  pages={537--574},
  year={2017},
  publisher={SIAM}
}

@article{kitsak2010identification,
  title={Identification of influential spreaders in complex networks},
  author={Kitsak, Maksim and Gallos, Lazaros K and Havlin, Shlomo and Liljeros, Fredrik and Muchnik, Lev and Stanley, H Eugene and Makse, Hern{\'a}n A},
  journal={Nature physics},
  volume={6},
  number={11},
  pages={888--893},
  year={2010},
  publisher={Nature Publishing Group UK London}
}

@article{garas2012k,
  title={A k-shell decomposition method for weighted networks},
  author={Garas, Antonios and Schweitzer, Frank and Havlin, Shlomo},
  journal={New Journal of Physics},
  volume={14},
  number={8},
  pages={083030},
  year={2012},
  publisher={IOP Publishing}
}

@article{li2013efficient,
  title={Efficient core maintenance in large dynamic graphs},
  author={Li, Rong-Hua and Yu, Jeffrey Xu and Mao, Rui},
  journal={IEEE transactions on knowledge and data engineering},
  volume={26},
  number={10},
  pages={2453--2465},
  year={2013},
  publisher={IEEE}
}

@article{sariyuce2013streaming,
  title={Streaming algorithms for k-core decomposition},
  author={Sar{\'\i}y{\"u}ce, Ahmet Erdem and Gedik, Bu{\u{g}}ra and Jacques-Silva, Gabriela and Wu, Kun-Lung and {\c{C}}ataly{\"u}rek, {\"U}mit V},
  journal={Proceedings of the VLDB Endowment},
  volume={6},
  number={6},
  pages={433--444},
  year={2013},
  publisher={VLDB Endowment}
}

@inproceedings{jakma2012distributed,
  title={Distributed k-core decomposition of dynamic graphs},
  author={Jakma, Paul and Orczyk, Marcin and Perkins, Colin S and Fayed, Marwan},
  booktitle={Proceedings of the 2012 ACM conference on CoNEXT student workshop},
  pages={39--40},
  year={2012}
}

@article{yin2018inter,
  title={Inter-layer similarity-based eigenvector centrality measures for temporal networks},
  author={Yin, Ran-Ran and Guo, Qiang and Yang, Jian-Nan and Liu, Jian-Guo},
  journal={Physica A: Statistical Mechanics and its Applications},
  volume={512},
  pages={165--173},
  year={2018},
  publisher={Elsevier}
}

@inproceedings{lee2012qube,
  title={Qube: a quick algorithm for updating betweenness centrality},
  author={Lee, Min-Joong and Lee, Jungmin and Park, Jaimie Yejean and Choi, Ryan Hyun and Chung, Chin-Wan},
  booktitle={Proceedings of the 21st international conference on World Wide Web},
  pages={351--360},
  year={2012}
}

@inproceedings{halatchliyski2010integrates,
  title={Who integrates the networks of knowledge in Wikipedia?},
  author={Halatchliyski, Iassen and Moskaliuk, Johannes and Kimmerle, Joachim and Cress, Ulrike},
  booktitle={Proceedings of the 6th international symposium on wikis and open collaboration},
  pages={1--10},
  year={2010}
}

@article{brandes2003communicating,
  title={Communicating centrality in policy network drawings},
  author={Brandes, Ulrik and Kenis, Patrick and Wagner, Dorothea},
  journal={IEEE transactions on visualization and computer graphics},
  volume={9},
  number={2},
  pages={241--253},
  year={2003},
  publisher={IEEE}
}

@inproceedings{zhang2011closeness,
  title={Closeness centrality on BBS reply network},
  author={Zhang, Ke and Li, Hui and Qin, Lijuan and Wu, Min},
  booktitle={2011 International Conference of Information Technology, Computer Engineering and Management Sciences},
  volume={2},
  pages={80--82},
  year={2011},
  organization={IEEE}
}

@inproceedings{jin2011based,
  title={Based on analyzing closeness and authority for ranking expert in social network},
  author={Jin, Ling and Yoon, Jae Yeol and Kim, Young Hee and Kim, Ung Mo},
  booktitle={International Conference on Intelligent Computing},
  pages={277--283},
  year={2011},
  organization={Springer}
}

@inproceedings{niu2011dgccf,
  title={DGCCF: Data gathering based on closeness centrality forwarding in opportunistic mobile sensor networks},
  author={Niu, Jianwei and Dai, Bin and Guo, Jinkai and Tong, Chao},
  booktitle={2011 7th International Conference on Wireless Communications, Networking and Mobile Computing},
  pages={1--6},
  year={2011},
  organization={IEEE}
}

@article{brohl2019centrality,
  title={Centrality-based identification of important edges in complex networks},
  author={Br{\"o}hl, Timo and Lehnertz, Klaus},
  journal={Chaos: An Interdisciplinary Journal of Nonlinear Science},
  volume={29},
  number={3},
  year={2019},
  publisher={AIP Publishing}
}

@article{gleich2015pagerank,
  title={PageRank beyond the web},
  author={Gleich, David F},
  journal={siam REVIEW},
  volume={57},
  number={3},
  pages={321--363},
  year={2015},
  publisher={SIAM}
}

@inproceedings{batagelj1999partitioning,
  title={Partitioning approach to visualization of large graphs},
  author={Batagelj, Vladimir and Mrvar, Andrej and Zaver{\v{s}}nik, Matja{\v{z}}},
  booktitle={International Symposium on Graph Drawing},
  pages={90--97},
  year={1999},
  organization={Springer}
}

@article{alvarez2005large,
  title={Large scale networks fingerprinting and visualization using the k-core decomposition},
  author={Alvarez-Hamelin, J and Dall'Asta, Luca and Barrat, Alain and Vespignani, Alessandro},
  journal={Advances in neural information processing systems},
  volume={18},
  year={2005}
}

@article{bader2003automated,
  title={An automated method for finding molecular complexes in large protein interaction networks},
  author={Bader, Gary D and Hogue, Christopher WV},
  journal={BMC bioinformatics},
  volume={4},
  number={1},
  pages={2},
  year={2003},
  publisher={Springer}
}

@article{altaf2003prediction,
  title={Prediction of protein functions based on k-cores of protein-protein interaction networks and amino acid sequences},
  author={Altaf-Ul-Amine, Md and Nishikata, Kensaku and Korna, Toshihiro and Miyasato, Teppei and Shinbo, Yoko and Arifuzzaman, Md and Wada, Chieko and Maeda, Maki and Oshima, Taku and Mori, Hirotada and others},
  journal={Genome Informatics},
  volume={14},
  pages={498--499},
  year={2003},
  publisher={Japanese Society for Bioinformatics}
}

@article{wuchty2005peeling,
  title={Peeling the yeast protein network},
  author={Wuchty, Stefan and Almaas, Eivind},
  journal={Proteomics},
  volume={5},
  number={2},
  pages={444--449},
  year={2005},
  publisher={Wiley Online Library}
}

@article{kim2012temporal,
  title={Temporal node centrality in complex networks},
  author={Kim, Hyoungshick and Anderson, Ross},
  journal={Physical Review E—Statistical, Nonlinear, and Soft Matter Physics},
  volume={85},
  number={2},
  pages={026107},
  year={2012},
  publisher={APS}
}

@article{elmezain2021temporal,
  title={Temporal degree-degree and closeness-closeness: A new centrality metrics for social network analysis},
  author={Elmezain, Mahmoud and Othman, Ebtesam A and Ibrahim, Hani M},
  journal={Mathematics},
  volume={9},
  number={22},
  pages={2850},
  year={2021},
  publisher={MDPI}
}

@incollection{nicosia2013graph,
  title={Graph metrics for temporal networks},
  author={Nicosia, Vincenzo and Tang, John and Mascolo, Cecilia and Musolesi, Mirco and Russo, Giovanni and Latora, Vito},
  booktitle={Temporal networks},
  pages={15--40},
  year={2013},
  publisher={Springer}
}

@article{galimberti2020span,
  title={Span-core decomposition for temporal networks: Algorithms and applications},
  author={Galimberti, Edoardo and Ciaperoni, Martino and Barrat, Alain and Bonchi, Francesco and Cattuto, Ciro and Gullo, Francesco},
  journal={ACM Transactions on Knowledge Discovery from Data (TKDD)},
  volume={15},
  number={1},
  pages={1--44},
  year={2020},
  publisher={ACM New York, NY, USA}
}

@article{oettershagen2022computing,
  title={Computing top-k temporal closeness in temporal networks},
  author={Oettershagen, Lutz and Mutzel, Petra},
  journal={Knowledge and Information Systems},
  volume={64},
  number={2},
  pages={507--535},
  year={2022},
  publisher={Springer}
}

@article{huang2017dynamic,
  title={Dynamic-Sensitive centrality of nodes in temporal networks},
  author={Huang, Da-Wen and Yu, Zu-Guo},
  journal={Scientific reports},
  volume={7},
  number={1},
  pages={41454},
  year={2017},
  publisher={Nature Publishing Group UK London}
}

@article{taylor1904tunable,
  title={Tunable eigenvector-based centralities for multiplex and temporal networks (2019)},
  author={Taylor, D and Porter, MA and Mucha, PJ},
  journal={arXiv preprint arXiv:1904.02059}
}

@article{flores2018eigenvector,
  title={On eigenvector-like centralities for temporal networks: Discrete vs. continuous time scales},
  author={Flores, Julio and Romance, Miguel},
  journal={Journal of Computational and Applied Mathematics},
  volume={330},
  pages={1041--1051},
  year={2018},
  publisher={Elsevier}
}

@inproceedings{rozenshtein2016temporal,
  title={Temporal pagerank},
  author={Rozenshtein, Polina and Gionis, Aristides},
  booktitle={Joint European conference on machine learning and knowledge discovery in databases},
  pages={674--689},
  year={2016},
  organization={Springer}
}

@article{aleja2024time,
  title={Time-dependent personalized PageRank for temporal networks: Discrete and continuous scales},
  author={Aleja, David and Flores, Julio and Primo, Eva and Romance, Miguel},
  journal={Chaos: An Interdisciplinary Journal of Nonlinear Science},
  volume={34},
  number={8},
  year={2024},
  publisher={AIP Publishing}
}

@inproceedings{cruciani2024mantra,
  title={Mantra: Temporal betweenness centrality approximation through sampling},
  author={Cruciani, Antonio},
  booktitle={Joint European Conference on Machine Learning and Knowledge Discovery in Databases},
  pages={125--143},
  year={2024},
  organization={Springer}
}

@article{barabasi2005origin,
  title={The origin of bursts and heavy tails in human dynamics},
  author={Barabasi, Albert-Laszlo},
  journal={Nature},
  volume={435},
  number={7039},
  pages={207--211},
  year={2005},
  publisher={Nature Publishing Group UK London}
}

@article{stehle2010dynamical,
  title={Dynamical and bursty interactions in social networks},
  author={Stehl{\'e}, Juliette and Barrat, Alain and Bianconi, Ginestra},
  journal={Physical Review E—Statistical, Nonlinear, and Soft Matter Physics},
  volume={81},
  number={3},
  pages={035101},
  year={2010},
  publisher={APS}
}

@article{takaguchi2016coverage,
  title={Coverage centralities for temporal networks},
  author={Takaguchi, Taro and Yano, Yosuke and Yoshida, Yuichi},
  journal={The European Physical Journal B},
  volume={89},
  number={2},
  pages={35},
  year={2016},
  publisher={Springer}
}

@article{rocha2014random,
  title={Random walk centrality for temporal networks},
  author={Rocha, Luis EC and Masuda, Naoki},
  journal={New Journal of Physics},
  volume={16},
  number={6},
  pages={063023},
  year={2014},
  publisher={IOP Publishing}
}

@article{lu2016vital,
  title={Vital nodes identification in complex networks},
  author={L{\"u}, Linyuan and Chen, Duanbing and Ren, Xiao-Long and Zhang, Qian-Ming and Zhang, Yi-Cheng and Zhou, Tao},
  journal={Physics reports},
  volume={650},
  pages={1--63},
  year={2016},
  publisher={Elsevier}
}

@article{bonacich1972factoring,
  title={Factoring and weighting approaches to status scores and clique identification},
  author={Bonacich, Phillip},
  journal={Journal of mathematical sociology},
  volume={2},
  number={1},
  pages={113--120},
  year={1972},
  publisher={Taylor \& Francis}
}

@article{liu2016locating,
  title={Locating influential nodes via dynamics-sensitive centrality},
  author={Liu, Jian-Guo and Lin, Jian-Hong and Guo, Qiang and Zhou, Tao},
  journal={Scientific reports},
  volume={6},
  number={1},
  pages={21380},
  year={2016},
  publisher={Nature Publishing Group UK London}
}

@article{boccaletti2006complex,
  title={Complex networks: Structure and dynamics},
  author={Boccaletti, Stefano and Latora, Vito and Moreno, Yamir and Chavez, Martin and Hwang, D-U},
  journal={Physics reports},
  volume={424},
  number={4-5},
  pages={175--308},
  year={2006},
  publisher={Elsevier}
}

@inproceedings{kempe2003maximizing,
  title={Maximizing the spread of influence through a social network},
  author={Kempe, David and Kleinberg, Jon and Tardos, {\'E}va},
  booktitle={Proceedings of the ninth ACM SIGKDD international conference on Knowledge discovery and data mining},
  pages={137--146},
  year={2003}
}

@article{holme2003congestion,
  title={Congestion and centrality in traffic flow on complex networks},
  author={Holme, Petter},
  journal={Advances in Complex Systems},
  volume={6},
  number={02},
  pages={163--176},
  year={2003},
  publisher={World Scientific}
}

@misc{brede2012networks,
  title={Networks—An Introduction. Mark EJ Newman.(2010, Oxford University Press.)},
  author={Brede, Markus},
  year={2012},
  publisher={MIT Press One Rogers Street, Cambridge, MA 02142-1209, USA journals-info~…}
}

@article{brin1998anatomy,
  title={The anatomy of a large-scale hypertextual web search engine},
  author={Brin, Sergey and Page, Lawrence},
  journal={Computer networks and ISDN systems},
  volume={30},
  number={1-7},
  pages={107--117},
  year={1998},
  publisher={Elsevier}
}

@article{langville2011google,
  title={Google's PageRank and beyond: The science of search engine rankings},
  author={Langville, Amy N and Meyer, Carl D},
  year={2011},
  publisher={Princeton university press}
}

@article{holme2015modern,
  title={Modern temporal network theory: a colloquium},
  author={Holme, Petter},
  journal={The European Physical Journal B},
  volume={88},
  number={9},
  pages={234},
  year={2015},
  publisher={Springer}
}

@book{saramaki2013temporal,
  title={Temporal networks},
  author={Saramaki, Jari and Holme, P},
  year={2013},
  publisher={Springer-verlag Berlin And Hei}
}

@article{wang2011identifying,
  title={Identifying and characterizing nodes important to community structure using the spectrum of the graph},
  author={Wang, Yang and Di, Zengru and Fan, Ying},
  journal={PloS one},
  volume={6},
  number={11},
  pages={e27418},
  year={2011},
  publisher={Public Library of Science San Francisco, USA}
}

@article{daley1965stochastic,
  title={Stochastic rumours},
  author={Daley, Daryl J and Kendall, David G},
  journal={IMA Journal of Applied Mathematics},
  volume={1},
  number={1},
  pages={42--55},
  year={1965},
  publisher={Oxford University Press}
}

@article{bonacich1987power,
  title={Power and centrality: A family of measures},
  author={Bonacich, Phillip},
  journal={American journal of sociology},
  volume={92},
  number={5},
  pages={1170--1182},
  year={1987},
  publisher={University of Chicago Press}
}

@book{anderson1991infectious,
  title={Infectious diseases of humans: dynamics and control},
  author={Anderson, Roy M and May, Robert M},
  year={1991},
  publisher={Oxford university press}
}

@book{keeling2008modeling,
  title={Modeling infectious diseases in humans and animals},
  author={Keeling, Matt J and Rohani, Pejman},
  year={2008},
  publisher={Princeton university press}
}

@article{newman2004finding,
  title={Finding and evaluating community structure in networks},
  author={Newman, Mark EJ and Girvan, Michelle},
  journal={Physical review E},
  volume={69},
  number={2},
  pages={026113},
  year={2004},
  publisher={APS}
}

@inproceedings{berger2006framework,
  title={A framework for analysis of dynamic social networks},
  author={Berger-Wolf, Tanya Y and Saia, Jared},
  booktitle={Proceedings of the 12th ACM SIGKDD international conference on Knowledge discovery and data mining},
  pages={523--528},
  year={2006}
}

@inproceedings{baeza2002web,
  title={Web structure, dynamics and page quality},
  author={Baeza-Yates, Ricardo and Saint-Jean, Felipe and Castillo, Carlos},
  booktitle={International Symposium on String Processing and Information Retrieval},
  pages={117--130},
  year={2002},
  organization={Springer}
}

@article{berberich2005time,
  title={Time-aware authority ranking},
  author={Berberich, Klaus and Vazirgiannis, Michalis and Weikum, Gerhard},
  journal={Internet Mathematics},
  volume={2},
  number={3},
  pages={301--332},
  year={2005},
  publisher={Taylor \& Francis}
}

@inproceedings{dong2010towards,
  title={Towards recency ranking in web search},
  author={Dong, Anlei and Chang, Yi and Zheng, Zhaohui and Mishne, Gilad and Bai, Jing and Zhang, Ruiqiang and Buchner, Karolina and Liao, Ciya and Diaz, Fernando},
  booktitle={Proceedings of the third ACM international conference on Web search and data mining},
  pages={11--20},
  year={2010}
}

@inproceedings{yu2004temporal,
  title={On the temporal dimension of search},
  author={Yu, Philip S and Li, Xin and Liu, Bing},
  booktitle={Proceedings of the 13th international World Wide Web conference on Alternate track papers \& posters},
  pages={448--449},
  year={2004}
}

@article{lu2011small,
  title={The small world yields the most effective information spreading},
  author={L{\"u}, Linyuan and Chen, Duan-Bing and Zhou, Tao},
  journal={New Journal of Physics},
  volume={13},
  number={12},
  pages={123005},
  year={2011},
  publisher={IOP Publishing}
}

@article{medo2009adaptive,
  title={Adaptive model for recommendation of news},
  author={Medo, Mat{\'u}{\v{s}} and Zhang, Yi-Cheng and Zhou, Tao},
  journal={Europhysics Letters},
  volume={88},
  number={3},
  pages={38005},
  year={2009},
  publisher={IOP Publishing}
}

@article{jeong2001lethality,
  title={Lethality and centrality in protein networks},
  author={Jeong, Hawoong and Mason, Sean P and Barab{\'a}si, A-L and Oltvai, Zoltan N},
  journal={Nature},
  volume={411},
  number={6833},
  pages={41--42},
  year={2001},
  publisher={Nature Publishing Group UK London}
}

@article{sporns2007identification,
  title={Identification and classification of hubs in brain networks},
  author={Sporns, Olaf and Honey, Christopher J and K{\"o}tter, Rolf},
  journal={PloS one},
  volume={2},
  number={10},
  pages={e1049},
  year={2007},
  publisher={Public Library of Science San Francisco, USA}
}

@article{teng2016collective,
  title={Collective influence of multiple spreaders evaluated by tracing real information flow in large-scale social networks},
  author={Teng, Xian and Pei, Sen and Morone, Flaviano and Makse, Hern{\'a}n A},
  journal={Scientific reports},
  volume={6},
  number={1},
  pages={36043},
  year={2016},
  publisher={Nature Publishing Group UK London}
}

@article{sabidussi1966centrality,
  title={The centrality index of a graph},
  author={Sabidussi, Gert},
  journal={Psychometrika},
  volume={31},
  number={4},
  pages={581--603},
  year={1966},
  publisher={Springer-Verlag}
}

@article{gao2014ranking,
  title={Ranking the spreading ability of nodes in complex networks based on local structure},
  author={Gao, Shuai and Ma, Jun and Chen, Zhumin and Wang, Guanghui and Xing, Changming},
  journal={Physica A: Statistical Mechanics and its Applications},
  volume={403},
  pages={130--147},
  year={2014},
  publisher={Elsevier}
}

@article{bonacich2001eigenvector,
  title={Eigenvector-like measures of centrality for asymmetric relations},
  author={Bonacich, Phillip and Lloyd, Paulette},
  journal={Social networks},
  volume={23},
  number={3},
  pages={191--201},
  year={2001},
  publisher={Elsevier}
}

@book{masuda2016guide,
  title={A guide to temporal networks},
  author={Masuda, Naoki and Lambiotte, Renaud},
  year={2016},
  publisher={World Scientific}
}

\end{document}